\documentclass[aps,prl,reprint,superscriptaddress,amsmath,amssymb,a4paper,twocolumn,longbibliography]{revtex4-2}

\usepackage[utf8]{inputenc}
\usepackage{amsmath}
\usepackage{amssymb,amsfonts,latexsym}
\usepackage{bm}
\usepackage[mathcal]{euscript}
\usepackage{graphicx}
\usepackage{epsfig}
\usepackage{color}
\usepackage{pifont,wasysym,marvosym}
\usepackage{textcomp}
\usepackage{comment}
\usepackage{epstopdf}
\usepackage{gensymb}
\usepackage{dcolumn}
\usepackage{hyperref}
\usepackage{physics}
\usepackage{babel} 
\renewcommand{\d}{\partial}

\newcommand{\nocontentsline}[3]{}
\let\origcontentsline\addcontentsline
\newcommand\stoptoc{\let\addcontentsline\nocontentsline}
\newcommand\resumetoc{\let\addcontentsline\origcontentsline}

\usepackage{etoolbox}
\makeatletter
\patchcmd{\@ssect}{\addcontentsline{toc}{#1}{\protect\numberline{}#8}}{}{}{}
\makeatother

\usepackage{siunitx}

\newcommand{\myauthors}{%
Jonas Veenstra\textsuperscript{1}, Jack Binysh\textsuperscript{1}, Vito Seinen\textsuperscript{1}, Rutger Naber\textsuperscript{1}, Damien Robledo Poisson\textsuperscript{1}, Andres Hunt\textsuperscript{2}, Wim van Saarloos\textsuperscript{3}, Anton Souslov\textsuperscript{4}, Corentin Coulais\textsuperscript{1} \\[0.5em]
\textsuperscript{1}\textit{Institute of Physics, Universiteit van Amsterdam, 1098 XH Amsterdam, The Netherlands} \\
\textsuperscript{2}\textit{Instituut-Lorentz, Universiteit Leiden, 2300 RA Leiden , The Netherlands} \\
\textsuperscript{3}\textit{TCM Group, Cavendish Laboratory, JJ Thomson Avenue, Cambridge, CB3 0US, United Kingdom}
}

\begin{document}

\def\scititle{Wave coarsening drives time crystallization in active solids}
\title{\scititle}


\author{Jonas Veenstra}
\thanks{These authors contributed equally}
\affiliation{Institute of Physics, Universiteit van Amsterdam, 1098 XH Amsterdam, The Netherlands}
\author{Jack Binysh}
\thanks{These authors contributed equally}
\affiliation{Institute of Physics, Universiteit van Amsterdam, 1098 XH Amsterdam, The Netherlands}
\author{Vito Seinen}
\affiliation{Institute of Physics, Universiteit van Amsterdam, 1098 XH Amsterdam, The Netherlands}
\author{Rutger Naber}
\affiliation{Institute of Physics, Universiteit van Amsterdam, 1098 XH Amsterdam, The Netherlands}
\author{Damien Robledo Poisson}
\affiliation{Institute of Physics, Universiteit van Amsterdam, 1098 XH Amsterdam, The Netherlands}
\author{Andres Hunt}
\affiliation{Department of Precision and Microsystems Engineering, Delft University of Technology, 2628 CD Delft, The Netherlands}
\author{Wim van Saarloos}
\affiliation{Instituut-Lorentz, Universiteit Leiden, 2300 RA Leiden, The Netherlands}
\author{Anton Souslov}
\email{as3546@cam.ac.uk}
\affiliation{
TCM Group, Cavendish Laboratory, JJ Thomson Avenue, Cambridge, CB3 0US, United Kingdom}
\author{Corentin Coulais}
\email{coulais@uva.nl}
\affiliation{Institute of Physics, Universiteit van Amsterdam, 1098 XH Amsterdam, The Netherlands}

\begin{abstract}
When metals are magnetized, emulsions phase separate, or galaxies cluster, domain walls and patterns form and irremediably coarsen over time. Such coarsening is universally driven by diffusive relaxation toward equilibrium. 
Here, we discover an inertial counterpart---wave coarsening---in active elastic media, where vibrations emerge and spontaneously grow in wavelength, period, and amplitude, before a globally synchronized state called a time crystal forms.
We observe wave coarsening in one- and two-dimensional solids and capture its dynamical scaling. We further arrest the process by breaking momentum conservation and reveal a far-from-equilibrium nonlinear analogue to chiral topological edge modes.
Our work unveils the crucial role of symmetries in the formation of time crystals and opens avenues for the control of nonlinear vibrations in active materials.
\end{abstract}

\maketitle
A striking manifestation of nonlinear physics is when microscopic periodic dynamics synchronize across space to produce coherent oscillations. 
Examples range from Faraday waves~\cite{Faraday1831-li,FortLevitating} to parametrically pumped optomechanical lattices~\cite{Slim2025-tm} and trapped ion chains~\cite{Zhang2017DiscreteTC}---which exhibit subharmonic oscillations arising from spontaneously broken discrete time-translation symmetry. 
More recently,
analogous phenomena have been explored with broken continuous time-translation symmetry, such as fireflies lighting up in synchrony~\cite{Peleg_NatRevPhys2024,National_Geographic_undated-xk}, cold atomic gases oscillating under optical pumping~\cite{Kongkhambut2022ContinuousTC,Wu2024DissipativeTC}, as well as biological collectives~\cite{PeyretBiophysicaljournal2019,XuNatPhys2022,TanNature,Chao2024-vg,guEmergenceCollectiveOscillations2025} and metamaterials~\cite{baconnier2022selective,ZheludevNatPhys, Veenstra_Nature2025}---where self-sustained oscillations emerge without any periodic drive. These are called time crystals.

Time crystals are typically driven externally---for example, by accelerating a water tank to make Faraday waves~\cite{Faraday1831-li}, or forcing atoms, ions or resonators using light~\cite{Zhang2017DiscreteTC,Kongkhambut2022ContinuousTC,Wu2024DissipativeTC,ZheludevNatPhys,delicnonreciprocity}. This external drive injects both energy and, crucially, momentum. An exception is active matter, which is energized from within. Active matter with elastic interactions has recently been shown to host time-crystalline phases~\cite{PeyretBiophysicaljournal2019, XuNatPhys2022,TanNature,guEmergenceCollectiveOscillations2025,baconnier2022selective,Veenstra_Nature2025}.
In this context, a key 
paradigm is non-reciprocal elasticity~\cite{Scheibner_NatPhys2020, Fruchart2023-uj}, in which energy is injected through an elastic tensor which is not symmetric
~\cite{TanNature,Bililign_NatPhys2021,Poncet_PRL2022,Guillet2025-xm,Chen_NatComm2021,Veenstra_Nature2025,Binysh2025-xf}. 
Unlike traditional routes to time crystallization, these odd elastic solids inject energy whilst respecting momentum conservation.
Here we ask: what is the role of conservation laws in the formation of time crystals?

We construct odd elastic metamaterials and discover a time-crystallization phenomenon that we dub \emph{wave coarsening}: active elastic waves spontaneously emerge, and their wavelength, period, and amplitude all grow in a self-similar fashion. This process arises from a nonlinear coupling between modes, sustained by active internal stresses. These stresses are generated through the injection of energy at the material’s microscopic scale and cascade to ever larger scales. 
Wave coarsening fundamentally differs from the traditional synchronization towards time crystals via phase locking. It is also distinct from traditional coarsening phenomena that relax towards equilibrium via diffusion.
Our results open new questions about the transient physics of systems with non-potential interactions~\cite{Fruchart2021non, ZheludevNatPhys,deWit_Nature_2024, ZheludevPRL2024} and suggest an organizing principle for nonlinear waves in acoustics~\cite{NoiraySynchronization,JaegerReconfiguration}, optomechanics~\cite{DelPino_Nature2022, Wanjura_NatPhys2023}, living matter~\cite{TanNature,VincenzoChirality}, and soft robotics~\cite{Veenstra_Nature2025} in the presence of conservation laws~\cite{braunsNonreciprocalPatternFormation2024,Greve2025-nh}.

\begin{figure*}[t!]
\begin{center}
\includegraphics[width=2.\columnwidth,trim=0cm 0cm 0cm 1cm]{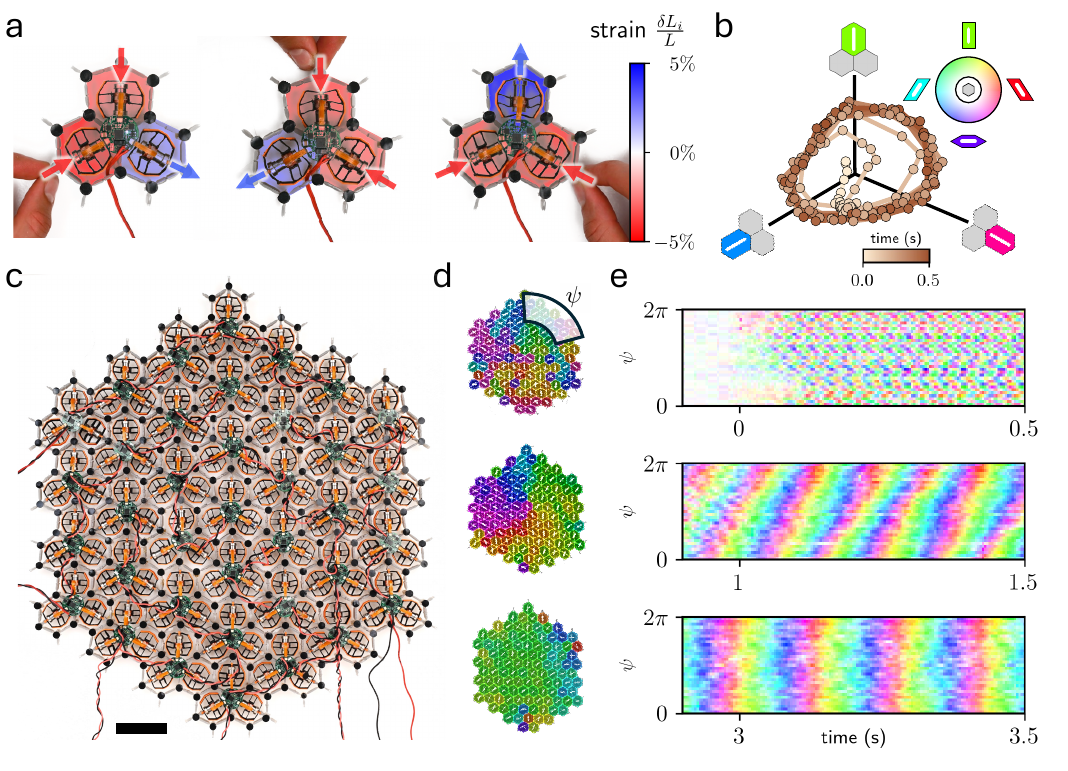}
\end{center}
\caption{\linespread{1.1}\selectfont{}
{\bf Wave coarsening in a non-reciprocal 2D metamaterial.} 
{\bf (a)} Non-reciprocal interactions between three hexagonal cells embedded with voice coil actuators, Hall sensors and a microcontroller.
When a hexagon is compressed (extended) along the actuator axis, its clockwise neighbor hexagon also compresses (extends) while its anticlockwise neighbor extends (compresses). This same response occurs regardless of the input cell.
{\bf (b)} Limit cycle dynamics in the unit cell beyond the threshold of stability. The axes correspond to the shear strains parallel to the actuator of each hexagon.
{\bf (c)} An odd elastic metamaterial consisting of 37 unit cells.
{\bf (d)} When the non-reciprocal interactions are driven beyond the threshold of instability of the material, high wavenumber shear modes are excited which move unidirectionally along the azimuthal axis of the material. Over time, the excitations gradually coarsen until the entire material shears synchronously. Data show the shear amplitudes $|\phi|$ and orientations $\arg\phi$ averaged over a radial window the size of which is equal to the instantaneous wavelength as a function of the azimuthal angle $\psi$ at times $t=[0.5,1.5,3.5]$s.
}
\label{COARSENING:Fig1}
\end{figure*}

Our active solids are mechatronic honeycombs (Fig.~\ref{COARSENING:trihex}). 
Each unit cell is made from a trio of elastic hexagons, 
whose strains are communicated locally to a custom printed circuit board which implements a nonreciprocal and non-pairwise relation between the deformations sensed and the forces actuated (SM~\S\ref{COARSENING:experimentalmethods}). 
The active forces $F_i$ on each site $i$ depend on the difference in deformation $\delta L_j - \delta L_k$ between two neighboring sites $j$ and $k$ via
\begin{align} 
\mqty( F_1\\F_2\\F_3 ) = 
\mqty(0 & k^a & - k^a\\ -k^a &0& k^a \\ k^a & -k^a &0)
\mqty(  \delta L_1\\ \delta L_2\\ \delta L_3  ),
\label{trihexDynMat}
\end{align} 
where $k^a$ sets the strength of the non-reciprocal interaction.
As a consequence, the unit cell breaks Maxwell-Betti reciprocity and responds asymmetrically regardless of which hexagon is compressed (Fig.~\ref{COARSENING:Fig1}a), a property not allowed in absence of non-potential forces~\cite{Scheibner_NatPhys2020}.

When $k^a$ is driven beyond a threshold value, the unit cell undergoes a Hopf bifurcation and exhibits a limit cycle (Fig.~\ref{COARSENING:Fig1}b). 
We now examine how this instability unfolds at the collective scale by arranging 37 unit cells into a 
honeycomb lattice (Fig.~\ref{COARSENING:Fig1}c).
With the material placed on top of a glass plate, lubricated with polytetrafluoroethylene (PTFE) to minimize friction, we now drive the nonreciprocal interactions beyond the bifurcation. 
Upon nudging the material out of equilibrium, we observe amplifying self-oscillations of around 30Hz at the level of each unit cell, which we measure and represent as a strain field in Fig.~\ref{COARSENING:Fig1}de. 
As the material oscillates, a longer wavelength mode slowly amplifies and drowns the initial unit cell oscillations. 
This wave travels unidirectionally along the edge 
before the system eventually settles into a globally synchronized oscillation with stable frequency, characteristic of a time-crystalline state. 
We refer to this transient process which is marked by the simultaneous growth of wavelength, period and amplitude, as wave coarsening.

\begin{figure}[t!]
\begin{center}
\includegraphics[width=1.\columnwidth,trim=0cm 0cm 0cm 0cm]{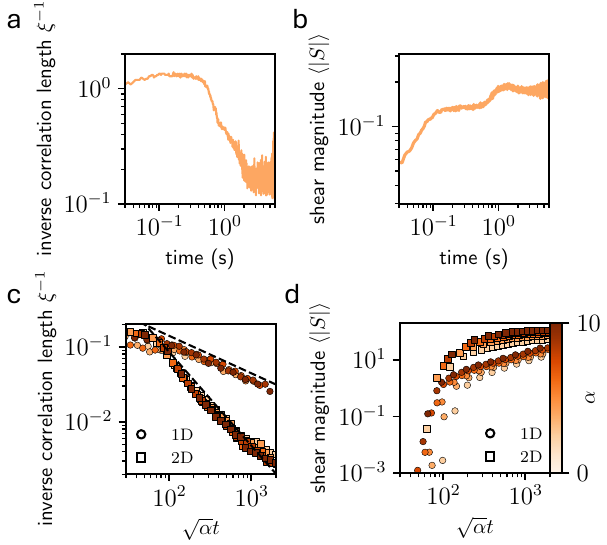}
\end{center}
\caption{\linespread{1.1}\selectfont{}
{\bf Spatial correlations suggest power-law coarsening.} 
 The inverse decay length $\xi^{-1}$ of the spatial correlation function {\bf (a)} and the average elastic energy $\langle|S|\rangle$ {\bf (b)}, plotted over time and averaged over six experiments.
{\bf (c)} The time evolution of $\xi^{-1}$ and elastic energy {\bf (d)} in numerical simulations of Eq.~\eqref{eq:phi_model} for both a 1D (round markers) and 2D grid (square markers) for a range of $\alpha$. Dashed lines on top of the 1D (2D) data correspond to a $-1/2$ ($-5/4$) power law.
}
\label{COARSENING:Fig2}
\end{figure}

To quantify the evolution of strain patterns, we track the inverse correlation length $\xi^{-1}$ of strains 
over time.
Averaged over space and across multiple runs, these data confirm that strain waves coarsen in wavelength over time (Fig. \ref{COARSENING:Fig2}a). 
Moreover, the rate at which the process unfolds suggests a power-law behavior. However, in tandem with the growth of wavelength, the strain amplitudes also steadily grow (Fig. \ref{COARSENING:Fig2}b), in contrast to gradient flow domain coarsening and synchronization phenomena.

To gain insight into the coarsening mechanism and its scaling behavior, we coarse-grain microscopic displacements at the unit cell level into a continuum field (Fig.~\ref{COARSENING:TriHexSchematic}, SM~\S\ref{Coarsening:coarsegrainingTrihex}):
\begin{equation}
\ddot{\phi} = \nabla^2 \left[ (1-i\alpha)\phi+ \dot{\phi} + |\phi|^2 \phi \right].
\label{eq:phi_model}
\end{equation}
Equation~\eqref{eq:phi_model} describes the nonlinear vibrations of a dissipative medium that is powered by odd elasticity, where shear modes couple antisymmetrically. The complex field $\phi(x,t)$ is the elastic strain field, whose real and imaginary parts $S_1$ and $S_2$ correspond to the two linearly independent shear modes. The parameter $\alpha$ is the non-dimensional activity that relates to the microscopic non-reciprocity $k^a$ and hinge stiffness $\kappa$ of our plaquettes as $\alpha = \sqrt{3}/8 k^a/\kappa $. Because energy is injected whilst conserving linear and angular momentum, the material does not undergo uniform translations or rotations but instead deforms. 

These conservation laws are encoded in the fully Laplacian form of Eq.~\eqref{eq:phi_model}, whose dispersion features gapless modes of wavenumber $q$ with complex frequencies $\sigma \approx \frac{1}{2}q(\alpha-q) + iq$ (Figs.~\ref{fig:PhiDispersion}~\ref{fig:SecondaryStability}, SM~\S\ref{COARSENING:PhiInstabilities}). All modes oscillate at a frequency proportional to their wavenumber $q$, while the non-reciprocal term amplifies modes with $q < \alpha$, with maximal growth at $q^* = \alpha/2$. Amplification at $q^*$ aligns with our observation that in the initial stage of wave coarsening traveling patterns with a characteristic wavenumber emerge before nonlinear effects saturate growth.

To confirm that waves coarsen in our continuum model, we investigate its nonlinear dynamics by numerically integrating Eq.~\eqref{eq:phi_model} with random initial conditions, on both one-dimensional and two-dimensional periodic domains. These simulations show that coherent wave patterns rapidly emerge and steadily coarsen in wavelength (Fig.~\ref{COARSENING:Fig2}c) with a characteristic power-law scaling while amplitudes steadily grow (Fig.~\ref{COARSENING:Fig2}d), regardless of the dimension (Fig.~\ref{fig:Coarsening1}), mirroring our experimental results.

Our results suggest that wave coarsening is a robust route to time crystallization, powered by momentum conserving energy injection. To test this hypothesis, we now consider a distinct non-reciprocal platform with a different dispersion and non-linearity.

\begin{figure*}[t!]
\begin{center}
\includegraphics[width=2.\columnwidth,trim=0cm 0cm 0cm 0cm]{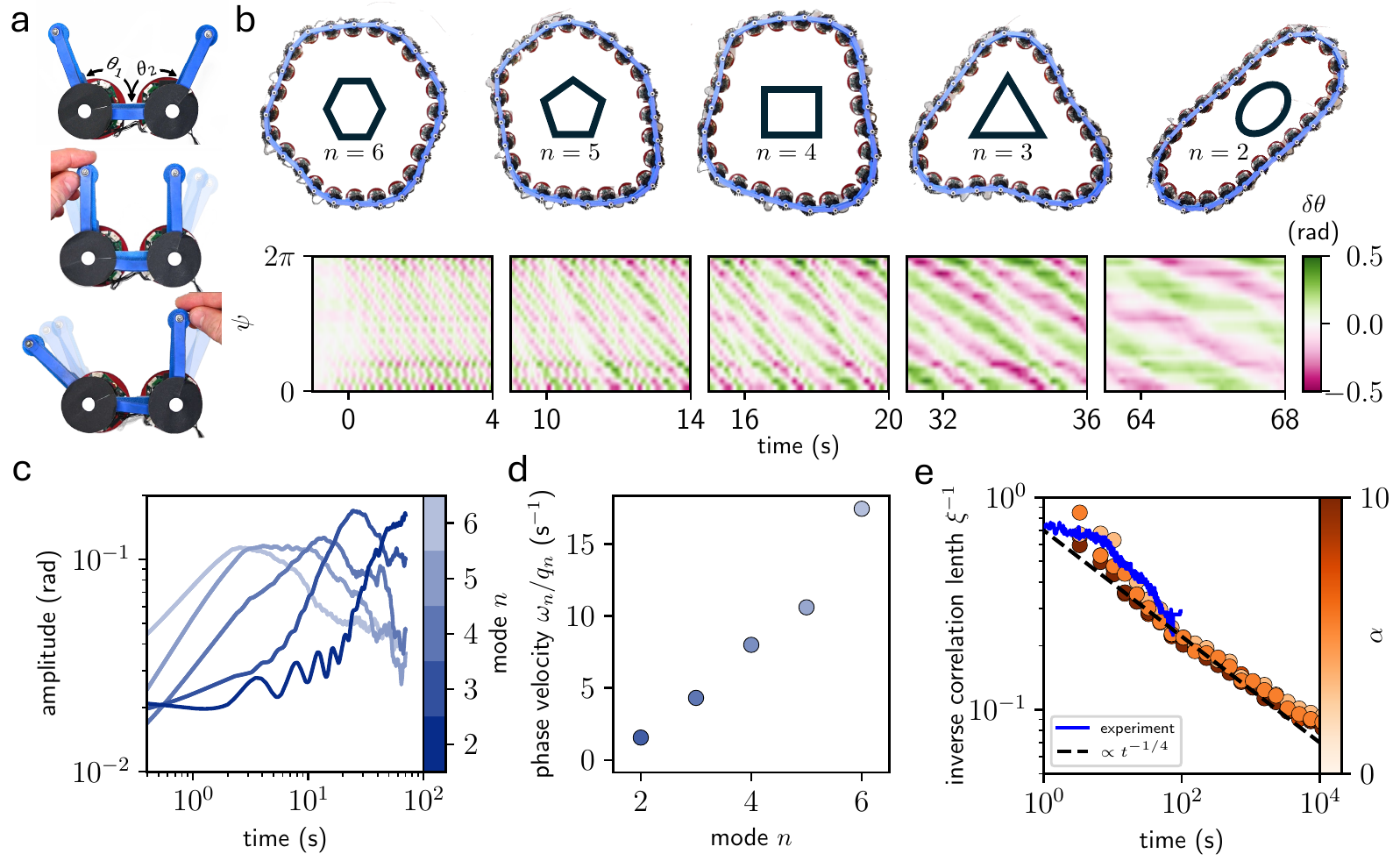}
\end{center}
\caption{\linespread{1.1}\selectfont{}
{\bf Wave coarsening in a 1D elastic chain with non-reciprocal interactions.} 
{\bf (a)} Non-reciprocal interaction between torsional unit cells. When angle $\theta_1$ is decreased, angle $\theta_2$ follows (middle). However, when angle $\theta_2$ is decreased, angle $\theta_1$ increases (bottom). 
{\bf (b)} A chain of 24 units undergoes an instability when the non-reciprocal coupling is driven beyond the Hopf bifurcation. This results in unidirectionally moving waves with a high wavenumber that gradually coarsen over time.
{\bf (c)} The mode amplitude over time for the lowest five modes.
{\bf (d)} Phase velocity of each mode.
{\bf (e)} The inverse correlation length over time in experiment (blue line) and simulations (markers) of Eq.~\eqref{eq:beam} for a range of the non-reciprocity $\alpha$. The dashed line indicates a power law of $\xi^{-1}\propto t^{-1/4}$.
}
\label{COARSENING:Fig3}
\end{figure*}

We construct a 1D periodic metamaterial in the form of an elastic ring embedded with 24 torque motors that respond antisymmetrically to angular deviations in neighboring sites 
(Fig.~\ref{COARSENING:Fig3}a, SM \S\ref{COARSENING:fibre}).
We place the ring upon an air table to let it float frictionlessly and drive the non-reciprocal coupling strength 
beyond its threshold of instability. 
Any small perturbation renders the ring unstable, causing waves with high mode number $n$ to emerge and grow in amplitude until eventually the motors saturate and deviate from their linearly non-reciprocal response. 
These waves then coalesce into longer wavelengths, deforming the ring into a series of unidirectionally morphing polygonal shapes with decreasing phase velocity and increasing amplitude (Fig. \ref{COARSENING:Fig3}b).

Like the two-dimensional metamaterials described above, wavelength, period and amplitude all increase during the coarsening process. Yet, in our one-dimensional ring, the time evolution of the mode structure is clearer (Fig. \ref{COARSENING:Fig3}c). 
While at early times, the $n=6$ mode quickly amplifies and briefly dominates the spectrum, soon the next highest mode $n=5$ catches up and overtakes. This victory for $n$th mode is again short-lived as it rapidly decays once its lower wavenumber neighbor $n-1$ takes over. As mode amplitudes transfer and mix, a travelling phase slip can be observed (Fig. \ref{COARSENING:Fig3}b). 
Eventually, the ring finds its dynamical steady state on a slow cycle around the lowest accessible deformation mode in the shape of a pancake. 

During the coarsening process, the wave velocity $\omega_n/q_n$, where $q_n=2\pi n /L$, consistently decreases
proportionally to
the mode number $n$ (Fig.~\ref{COARSENING:Fig3}d). This suggests the existence of a quadratic dispersion characteristic of flexural waves that is still valid in the nonlinear regime and coincides with a characteristic amplitude for each wavelength.
Finally, 
the inverse correlation length $\xi^{-1}$ plotted over time reveals another power-law trend, which we fit at $\xi^{-1}\propto t^{-0.29}$ (Fig.~\ref{COARSENING:Fig3}e). 

In order to consolidate these findings, we derive a coarse-grained description of a 1D non-reciprocal ring with a saturating nonlinearity, describing its real-valued deviations from its equilibrium state as a height displacement field $h(x,t)$ (SM~\S\ref{COARSENING:fibre}):
\begin{equation}
\ddot{h} =- \partial^2_x \left[ \partial^2_x h-\alpha\Bigg(\frac{\partial^3_x h}{\sqrt{1+(\partial^3_x h)^2}}\Bigg)+ \partial^2_x \dot{h}\right].
\label{eq:beam}
\end{equation}
Analogously to the 2D model Eq.~\eqref{eq:phi_model}, the right hand side of Eq.~\eqref{eq:beam} is entirely inside the Laplacian and thus conserves momentum whilst breaking reciprocity.
However, in this beam equation, the flexural deformations are inherently dispersive, and the nonlinearity takes a different form. 
To test whether coarsening persists under these conditions, we simulate Eq.~\eqref{eq:beam} and observe a clear power-law growth of the inverse correlation length $\xi^{-1}$, with an exponent close to $-1/4$, which is consistent with our experimental findings (Fig. \ref{COARSENING:Fig3}e). 
In summary, despite the differences in dimension, dispersion and non-linearity, waves in our 1D non-reciprocal ring grow in both wavelength and amplitude in a similar fashion to the 2D odd elastic metamaterial of Fig.~\ref{COARSENING:Fig1}, yet with a different dynamical exponent.

The data of Fig.~\ref{COARSENING:Fig3} strongly suggests a picture of various modes interacting with one another but the precise nature of this coupling remains unclear. To gain analytical insight, we write our continuum theories of Eqs.~\eqref{eq:phi_model},~\eqref{eq:beam} in the Fourier domain and make a mean field approximation for the interaction between modes (Figs.~\ref{COARSENING:fig_neglecting},~\ref{fig:NonlinearGrowthrate},~\ref{fig:MultipleScales}). Further averaging over fast oscillations in favor of the slow coarsening dynamics allows us to derive an evolution equation for the mode amplitudes which remains valid at the nonlinear level:
\begin{equation}
    \frac{\dot{|a_n|}}{|a_n|} = \frac{1}{2} \left( \frac{ \alpha }{ \sqrt{1 + U(t)} }q_n^p - {q_n^{2p}} \right).
\label{eq:NonlinearGrowthRateFirstOrder}
\end{equation}
Here $|a_n|$ and $\dot{|a_n|}$ denote the mode amplitudes and their growth rate, $p=1$ for the odd elastic system of Eq.~\eqref{eq:phi_model} and $p=2$ for the non-reciprocal beam of Eq.~\eqref{eq:beam}. Crucially, $U(t)=2\sum_m {q_m}^{6(p-1)}|a_m|^2$ represents an amplitude-dependent rescaling of the activity $\alpha$ that invites a compelling picture.

At the start of the coarsening process, modes grow according to their linear dispersion, and the most unstable wavelength is selected by $\alpha$.
However, as modes inevitably amplify, eventually the nonlinear term $U(t)$ 
grows to become comparable to the linear terms, which shifts the peak of the dispersion to a lower wavenumber and growth rate simultaneously (Fig.~\ref{fig:NonlinearGrowthrate}).
This new wavenumber now grows, albeit at a slower rate, and further increases $U(t)$, in turn pushing the dominant wavenumber and growth rate even lower.
A feedback loop ensues until a finite system size prevents a further cascade of wavenumbers. The resulting progressive shift in the dispersion curve suggests a self-similar dynamics: as energy builds up in the system, non-reciprocal forces drive growth at ever-larger length scales and at ever-slower rates.

To understand the timescales at which wave coarsening occurs, we search for solutions to Eq.~\eqref{eq:NonlinearGrowthRateFirstOrder}. Motivated by the experimental observation that even beyond the linear instability, waves remain quite coherent with a narrowly distributed mode structure, we use as as ansatz a wavepacket whose central wavenumber is larger than its width~(SM~\S\ref{COARSENING:AmplitudeEquations}).
This yields an evolution equation for the dominant length scale. Then, using a power-law ansatz $\xi^{-1}\propto t^{\beta}$ for the inverse correlation length, we find power laws $\beta=-1/(2p)=-1/4$ for Eq.~\eqref{eq:beam}, confirming the experimental and numerical results of Fig.~\ref{COARSENING:Fig3}e. In the 1D case of the nonlinearly odd elastic model of Eq.~\ref{eq:phi_model}, we similarly find $\beta=-1/(2p)=-1/2$, which is in agreement with numerical data (Fig.~\ref{COARSENING:Fig2}c). It is remarkable that our mean-field picture successfully captures the coarsening dynamics despite neglecting the details of the phase slip events. However, our approximation breaks down in 2D, where faster coarsening occurs through a sequence of strain defect pair annihilation events. We interpret the difference in coarsening rates
as follows. 
While in 1D, wave coarsening is driven by mode competition as described by Eq.~\eqref{eq:NonlinearGrowthRateFirstOrder}, with phase slips occurring only sporadically, the 2D counterpart is dominated the dynamics of mobile and attracting defects, which accelerates coarsening.

\begin{figure*}[t!]
\begin{center}
\includegraphics[width=1\linewidth,trim=0cm 0cm 0cm 0cm]{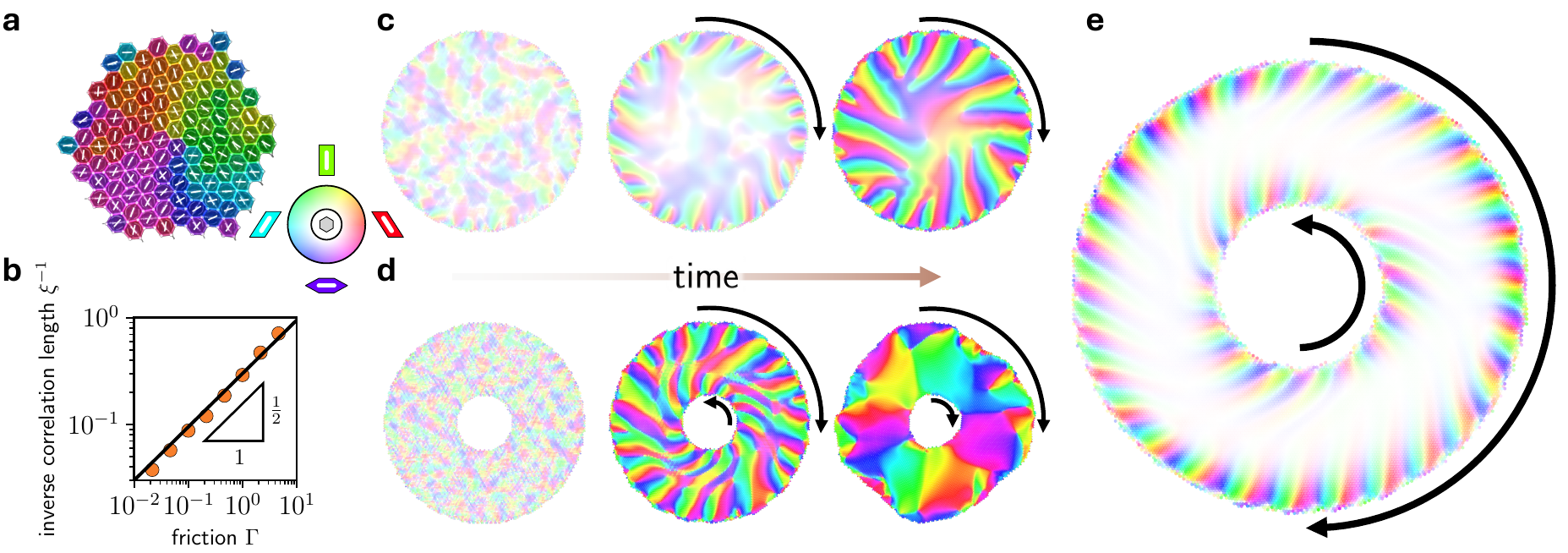}
\end{center}
\caption{\linespread{1.1}\selectfont{}
{\bf Frictional arrest of wave coarsening and edge waves.} 
{\bf (a)} A snapshot of the 2D metamaterial atop an unlubricated surface during wave coarsening demonstrates a stable defect in the phase of the strain field $\arg \phi$.
{\bf (b)} The steady state inverse correlation length $\xi^{-1}$ as a function of the friction $\Gamma$, found from numerical simulations of Eq.~\eqref{eq:NonlinearGrowthRateFirstOrder}. {\bf (c,d)}
Snapshots of a large-scale mass-spring simulation of the 2D metamaterial of Fig.~\ref{COARSENING:Fig1} on a circular {\bf (c)} and an annular {\bf (d)} domain show the gradual invasion of edge waves as time passes.
{\bf (e)} The steady state of the same annular system in the presence of friction.
}
\label{COARSENING:Fig4}
\end{figure*}

Similar phase slips and defect annihilations occur in a plethora of coarsening and synchronizing systems~\cite{Bray2002-tn,Yurke1993-ak, politi_when_2004,suchanek_entropy_2023,
braunsNonreciprocalPatternFormation2024, Chardac2021-cp, ranaCoarseningTwodimensionalIncompressible2020}. 
Yet in these cases, the coarsening dynamics is effectively a gradient flow that minimizes a global potential. By contrast, here the patterns and defects observed are nonlinear excitations that can only be sustained by continuous energy injection. Instead of relaxing toward a fixed potential minimum, the system redistributes energy as the nonlinear dispersion relation progressively favors larger length scales.

Now that we have captured the essence of wave coarsening, we ask in the remainder of the paper how robust it is to perturbations. We consider two perturbations: (i) breaking of momentum conservation and (ii) edge effects. 

First, we break momentum conservation. Our mean field analysis showed that wave coarsening continues until the wavelengths become system-sized, ultimately due to the fact that the linear growth rate $\Re \sigma \propto q(\alpha -q) $ is positive for all wavenumbers $q<\alpha$ , a consequence of the momentum conserving nature of the energy injection. We now gap these low $q$ modes by breaking momentum conservation, which can be achieved by enforcing momentum exchange of the material with a substrate, i.e. through friction.

We now remove the PTFE lubrication between the 2D material of Fig.~\ref{COARSENING:Fig1} and the substrate and observe how wave coarsening is arrested indefinitely while the system cycles around higher mode solutions without cascading further (Fig.~\ref{COARSENING:Fig4}a). 
To quantify this behavior, we numerically integrate Eq.~\eqref{eq:phi_model} with an added friction term proportional to $\Gamma \dot{\phi}$, and find that the steady state correlation length $\xi^{-1}$ perfectly coincides with the marginally stable wavelength predicted by our stability analysis (Fig.~\ref{COARSENING:Fig4}b and Fig~\ref{fig:SecondaryStability}).

Second, we consider the effect of boundary conditions.
To this end, we employ large-scale discrete simulations of the 2D material. Starting with a disk consisting of $5813$ unit cells, these reveal that while coarsening initially proceeds homogeneously throughout the bulk, chiral edge waves soon appear and eventually dominate the dynamics. These are Rayleigh waves that are unidirectionally amplified by non-reciprocity and stabilised by nonlinearities (Fig.~\ref{fig:RayleighWaves}, SM~\S\ref{COARSENING:Rayleigh}). As their amplitude grows, so does their wavelength and penetration depth, causing them to progressively invade the bulk (Fig.~\ref{COARSENING:Fig4}c). At early times, coarsening is driven by the annihilation of free-roaming strain defects of opposite charge. Once edge waves emerge, however, they impose spatial order on the defects and confine them to be regularly spaced. Coarsening then proceeds through discrete events reminiscent of the phase slips seen in Fig.~\ref{COARSENING:Fig2}, as defects are gradually expelled at the system’s edge, each time increasing the edge wavelength. 

To confirm the crucial role of edges in our system, we repeat our simulations on an annular domain and observe edge waves appear on both the inner and outer edge and propagate in opposite directions (Fig.~\ref{COARSENING:Fig4}d). As their amplitude and wavelength grows, these edge waves eventually meet and the outer edge overwhelms the inner one due its larger circumference. Their counterpropagating nature is reminiscent of chiral edge states in Chern insulators~\cite{Haldane1988-aw, Khanikaev2015Topologically, Nash2015-fc,Souslov2017-il, Delplace2017Topological, Zhang2021Superior,Fossati2024-wq}. However, because these waves progressively invade the bulk we do not yet have an insulating state.

Inspired by this analogy we again include friction, drawing on the finding that it selects a steady-state length scale and arrests waves from coarsening. Since the penetration depth of Rayleigh waves is proportional to their wavelength (SM~\S\ref{COARSENING:Rayleigh}), we find that the system now maintains an insulating state with counter-propagating edges waves, representing a striking nonlinear, far-from-equilibrium counterpart to the quantum Hall effect (Fig.~\ref{COARSENING:Fig4}e). This demonstrates how wave coarsening irremediably occurs even in the presence of perturbations, but the final time-crystalline state is sensitive to both the breaking of momentum conservation and to boundary conditions.

In conclusion, we have uncovered a nonlinear feedback mechanism in active elastic media,
where momentum-conserving energy injection drives waves to ever larger wavelength, amplitude and period.
Looking ahead, we anticipate that wave coarsening represents a paradigmatic route towards time-crystallization in systems with additional symmetries and associated Goldstone modes, that can be leveraged as a control strategy for shape changes and directed motion in fields such as robotics, active matter and optomechanics with non-reciprocal phase transitions~\cite{Fruchart2021non,hanaiNonreciprocalFrustrationTime2024}.

\stoptoc
\bibliography{references}

\section*{Acknowledgments}
We thank Tjeerd Weijers, Kasper van Nieuwland, Sven Koot, Melanie Pichon, for technical assistance and Colin Scheibner, Michel Fruchart and Gustavo D\"uring for valuable discussions.
C.C. and J.V. acknowledge funding from the European Research Council under Grant Agreement No.~852587 and from the Netherlands Organisation for Scientific Research (NWO) under grant agreement VI.Vidi.213.131.3. J.B. acknowledges funding from the European Union’s Horizon research and innovation programme under the Marie Sklodowska-Curie Grant Agreement No. 101106500. This research was supported in part by grant NSF PHY-2309135 to the Kavli Institute for Theoretical Physics (KITP) and by the Dutch Institute for Emergent Phenomena (DIEP).

\clearpage
\onecolumngrid

\resumetoc
\clearpage

\begin{center}
\stoptoc
\section*{Supplementary Materials for
}
\resumetoc
\end{center}
\onecolumngrid
\begin{center}
  \textbf{\large \scititle}\\[0.75em]
  \myauthors
\end{center}
\vspace{1em}

\setcounter{section}{0}       
\setcounter{subsection}{0}    
\setcounter{subsubsection}{0} 
\setcounter{equation}{0}      
\setcounter{figure}{0}        
\setcounter{table}{0}         

\setcounter{secnumdepth}{1} 
\renewcommand{\theequation}{S\arabic{equation}}   
\renewcommand{\thefigure}{S\arabic{figure}}       

\tableofcontents

\clearpage

\section{Materials and methods}\label{COARSENING:experimentalmethods}

\subsection{Construction of the two-dimensional active metamaterial}

\begin{figure*}[t!]
\begin{center}
\includegraphics[width=1.\columnwidth,trim=0cm 0cm 0cm 0cm]{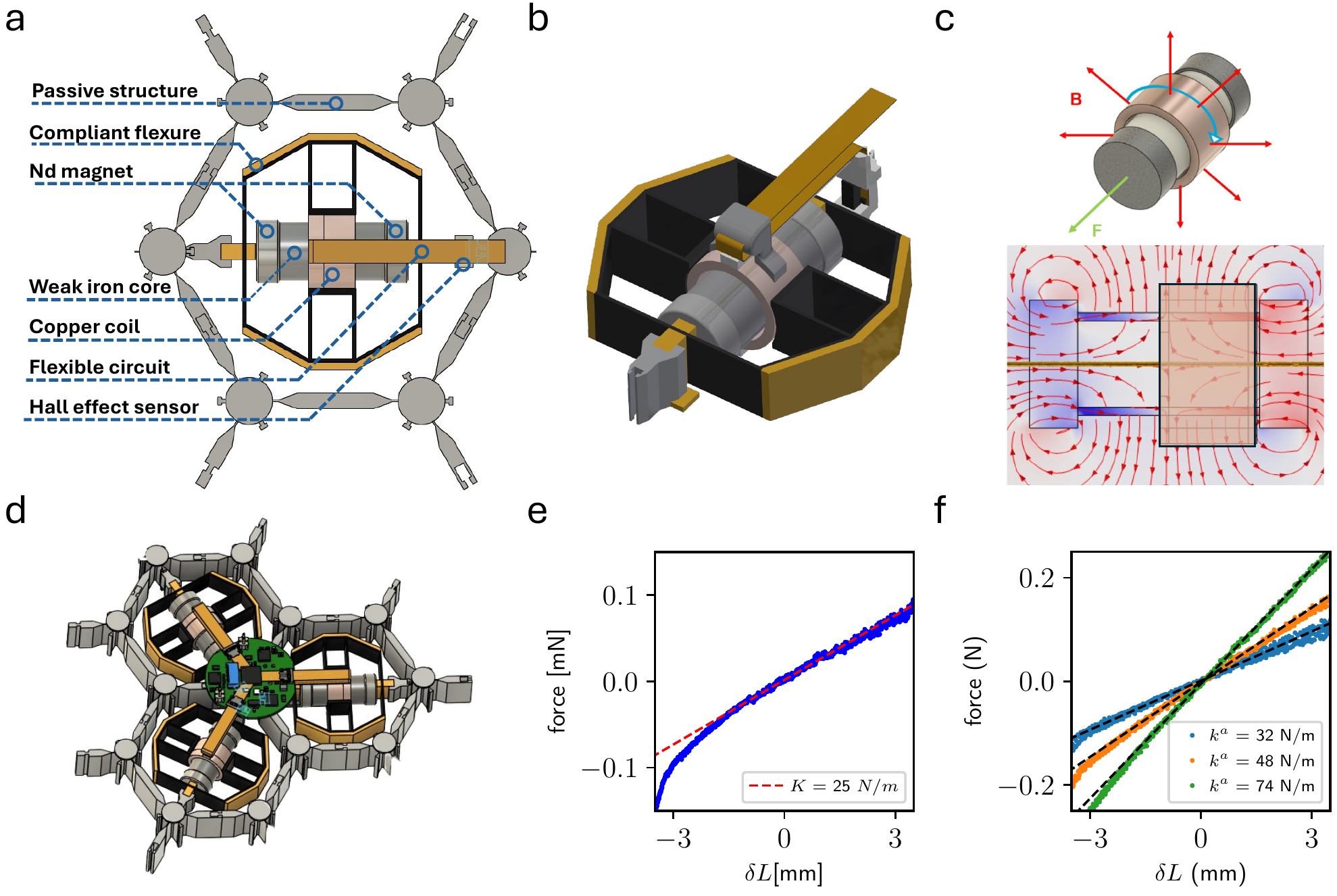}
\end{center}
\caption{\linespread{1.1}\selectfont{}
\textbf{2D odd elastic metamaterial with linear actuators.} 
{\bf (a)} A single passive structure (grey) with an embedded compliant flexure containing a voice coil actuator and a Hall effect sensor.
{\bf (b)} The compliant flexure consists of a stiff (orange) and a soft (black) part (PLA) and houses the integrated voice coil actuator and Hall sensor, seen from an isometric perspective. 
{\bf (c)} Top: the voice coil actuator, consisting of a weak iron core, sandwiched between two Neodymium magnets and surrounded by a copper coil, that freely translates along the core. Bottom: a longitudinal slice of the actuator showing magnetic field lines that penetrate the copper coil perpendicularly.
{\bf (d)} Three active hexagons communicate through a microcontroller sitting on top of the central vertex. Based on Hall effect sensor data, the microcontroller calculates strains in each hexagon and is programmed to exert the non-reciprocal force rule of Eq.~\eqref{trihexDynMat}. 
{\bf (e)} Force-displacement data characterizing the stiffness of the single unit cell depicted in panel (a).
{\bf (f)} Force-displacement data for the actuator coil as it responds to deformations of a neighboring unit cell. 
}
\label{COARSENING:trihex}
\end{figure*}

The metamaterial shown in Fig.~\ref{COARSENING:Fig1} consists of 3D-printed hexagonal structures (Formlabs Form 2 SLA printer with Tough1500 resin) with an embedded compliant flexure (E3D ToolChanger multi nozzle fused filament fabrication) that houses a custom-made voice coil actuator, as shown in Fig.~\ref{COARSENING:trihex}a. 
The actuator consists of a weak iron cylindrical core attached on both ends to neodymium magnets aligned to repel (Fig.~\ref{COARSENING:trihex}b). The resulting magnetic field lines meet in the middle and bend radially outwards perpendicularly to a copper coil that is wound around the weak iron core. The radius of the copper coil is larger than the radius of the iron core, which allows the coil to translate along the core (Fig.~\ref{COARSENING:trihex}c).

When an electrical current is turned on, a Lorentz force is generated that pushes the iron core in the opposite direction of the copper coil along the vertical direction. The compliant flexure ensures that the coil remains aligned with the core and transmits forces to the outer passive structure.
Attached to the right coil vertex is a Texas Instruments linear Hall effect sensor (DRV5057A3), which produces a voltage proportional to the magnetic field it senses. When the hexagon deforms along its actuator axis, the distance between the core and the Hall sensor changes, resulting in a voltage difference which we calibrate to the corresponding deformation of the hexagon.

We now connect sets of three hexagons together as shown in Fig.~\ref{COARSENING:trihex}d, with the actuators pointing symmetrically along the three principle axes of the hexagons which guarantees isotropy at the large scale. Each hexagon contains a flexible circuit that connects its Hall sensor and the terminals of its copper coil to a custom made printed circuit board (pcb), that sits on top of the central vertex. Integrated in the pcb is a 32-bit micro controller (ATSAMD21E18A) that controls the three bidirectional motor drivers, which can manipulate the current through the voice coil. The central circuit board collects Hall voltages from each hexagon and converts these signals into the corresponding deformations $\delta L = L_0 - L$ where $L_0=2l$ is the hexagon's long diagonal at equilibrium in terms of the hexagon side length $l=2$cm. The pcbs are powered by a 30A current source. We now implement non-reciprocal interactions between the forces exerted by the actuators and the shear deformations of neighboring hexagons according to the following antisymmetric dynamical matrix:
\begin{align} 
\mqty( F_1\\F_2\\F_3 ) = 
\mqty(0 & k^a & - k^a\\ -k^a &0& k^a \\ k^a & -k^a &0)
\mqty(  \delta L_1\\ \delta L_2\\ \delta L_3  ).
\label{trihexDynMat}
\end{align} 

Using a pulse-width modulated voltage, each hexagon coil receives a current which generates the non-reciprocal forces. Using an Instron tensile tester, we calibrate the passive stiffness of a hexagonal unit cell to be $K=25$N/m (Fig.~\ref{COARSENING:trihex}e) and find the actuator to exhibit a linear force in response to deformations $\delta L$ (Fig.~\ref{COARSENING:trihex}f).

\subsection{Construction of the one-dimensional active ring}
The 1D metamaterial shown in Fig.~\ref{COARSENING:Fig3}a is composed of motorised vertices connected by plastic arms. 
Each vertex consists of a DC coreless motor (Motraxx CL1628) embedded in a cylindrical heatsink, an angular encoder (CUI AMT113S), and a microcontroller (ESP32) connected to a custom electronic board. The electronic board enables power conversion, interfacing between the sensor and motor, and communication between vertices.
Each vertex has a diameter $50$mm, height $90$mm, and mass $0.2$kg. 
The power necessary to drive the motor is provided 
by an external 48V DC power source.

Rigid 3D-Printed arms connect each motor's drive shaft to the heat sink of the adjacent unit. The angle formed between the two arms at vertex $i$ is denoted by $\theta_i$. The on-board sensor measures $\theta_i$ at a sample rate of $100$Hz and communicates the measurement to nearest neighbours. In response to the incoming signal, vertex $i$ exerts an active torsional force $\tau_i^a$,
\begin{equation}
    \tau^a_{i} = \begin{cases}
        \kappa^a \left(\delta \theta_{i+1}-\delta \theta_{i-1} \right), & \mathrm{if}\,\,|\delta \theta_{i+1}-\delta \theta_{i-1} |<\tau_{\mathrm{max}}/\kappa^a \\
        \mathrm{sgn}(\delta \theta_{i+1}-\delta \theta_{i-1} )\tau_{\mathrm{max}}, & \mathrm{if}\,\,|\delta \theta_{i+1}-\delta \theta_{i-1} |>\tau_{\mathrm{max}}/\kappa^a,
    \end{cases}
\end{equation}
where $\delta \theta_i = \theta_i - \theta^0$, $\theta^0$ is the rest angle and $\tau_\mathrm{max}$ is the torque at which the motors saturate. The active stiffness $\kappa^a$ was programmed by the microcontroller and calibrated by measuring torque-displacement slopes for different values of the electronic feedback. 

The coreless motor saturates at a maximum torque of $\tau_\mathrm{max}=12$mN\,m. Each link has a length of $a=7.5\,\mathrm{cm}$. Adjacent vertices are also connected by a rubber band [blue in Fig.~\ref{COARSENING:Fig2}(a)] with a thickness 4mm which provides a passive elastic torsional stiffness of $\kappa=48$mN m/rad. 
For further details, see Ref.~\cite{Veenstra_Nature2025}. 

\subsection{2D unit cell}

\subsubsection{Two-dimensional active metamaterial} To observe deformations in the 2D metamaterial unhindered by frictional forces from the substrate on top of which it is placed (data shown in Fig.~\ref{COARSENING:Fig1} and Fig.~\ref{COARSENING:Fig2}), we attach small polytetrafluoroethylene (PTFE) cylinders on the underside of each hexagon. We then place the material on a glass plate to allow imaging from below and coat the glass plate with PTFE spray to minimize frictional effects. We track the position of the white PTFE supports from image data captured using a XIMEA CB209 camera at 10MP and 300fps, with an accuracy of $0.5$mm. The individual pcbs of all unit cells contain an integrated IR-sensor that we communicate with through a remote control which allows us to turn on and tune the non-reciprocal interactions on demand. We manually perturb the material upon which unit cells destabilize and strains amplify. We run a series of 6 repeated experiments, each of which fully coarsen within 6 seconds. In Fig.~\ref{COARSENING:Fig4}a, the PTFE coating was removed from the glass plate to increase friction.

We parameterize the internal deformation of hexagons via the three degrees of freedom
\begin{align}
    S_1 =& \frac1{\sqrt 2} ( \delta \theta_1 - \delta \theta_3 ), \\ 
    S_2=&  \frac1{\sqrt 6} (\delta \theta_1 + \delta \theta_3 - 2 \delta \theta_5), \\
    B=&  \frac1{\sqrt 3} (\delta \theta_1 + \delta \theta_3 + \delta \theta_5).
    \label{eq:Sprojection}
\end{align}
The two lowest modes of an incompressible hexagon are given by the shears $S_1$ and $S_2$. The additional breathing mode $B$ is allowed by the geometry of the hexagon but this deformation does not generate any odd forces and thus dissipates away. 
In Fig.~\ref{COARSENING:Fig1}b, we show shear strain data $\phi=S_1+iS_2$ projected onto the actuator direction of each of the three hexagons that together form a unit cell. 
We plot the shear state of each hexagon and overlay it on three snapshots corresponding to $t=[0.5,1.5,3.5]$ in Fig.~\ref{COARSENING:Fig1}d, and similarly for Fig.~\ref{COARSENING:Fig4}acde. In Fig.~\ref{COARSENING:Fig1}e, we project the shear data onto the azimuthal variable $\psi$ and plot it over over time. In Fig.~\ref{COARSENING:Fig2}, we plot the inverse of the decay length $\xi$ of the equal-time correlation $C(r)$, which we average over 6 experiments. 
The field $\phi(\mathbf{x}) = A(\mathbf{x}) e^{i\theta(\mathbf{x})}$ is complex, so we define the two-point correlator 
$C(r) = \langle A(\mathbf{x}) A(\mathbf{x}+\mathbf{r}) \cos[\theta(\mathbf{x}) - \theta(\mathbf{x}+\mathbf{r})] \rangle$, 
averaging over all pairs with separation $|\mathbf{r}|$ within bins of unit cell width in the case of experimental data (Fig.~\ref{COARSENING:Fig2}a) or spatial discretization in the case of simulation data (Fig.~\ref{COARSENING:Fig2}c). We normalize $C(r)$ such that $C(0) = 1$ and define the decay length $\xi$ as the smallest $r$ for which $C(r) < 1/e$.

\subsubsection{One-dimensional active ring}
The experimental results demonstrated in Fig.~\ref{COARSENING:Fig3} were obtained by tracking over time the vertices of an active chain consisting of 24 units suspended on airbearings. In order to set up randomized and small amplitude initial conditions, the ring was allowed to find equilibrium in its passive state, where the non-reciprocal torsional coupling $\kappa^a$ equals zero. At this point the state of the system was measured by the sensors and saved as a reference to calculate the active force rule. Then, still in its passive state, the system was manually excited and again allowed to equilibrate. Due to small frictional effects, this new physical equilibrium only slightly different from the reference state used by the motor feedback, but large enough to be picked up by the sensitive 10-bit angular decoder. As a result, this procedure successfully generates small amplitude random initial conditions.

To observe coarsening, we now drive up the non-reciprocal coupling $\kappa^a$ well beyond the Hopf bifurcation and destabilize one of the higher modes accessible to the ring. We use the data recorded by the angular decoders at a rate of $50$Hz and cross validate the data with observations of the motion of each vertex over time with a Nikon D5600 camera. Experiments run for $90$ seconds and we average over a dataset of $10$ experimental runs, yielding the data shown in Fig.~\ref{COARSENING:Fig3}cde. 
We define the inverse correlation length $\xi^{-1}$ as before, and find an experimental coarsening law $\xi^{-1} \propto t^{-0.29}$ by fitting the $\xi^{-1}$ versus time.

\subsection{Calculating odd moduli via coarse graining}\label{Coarsening:coarsegrainingTrihex}
\begin{figure*}[t]
\begin{center}
\includegraphics[width=1\linewidth,trim=0cm 0cm 0cm 0cm]{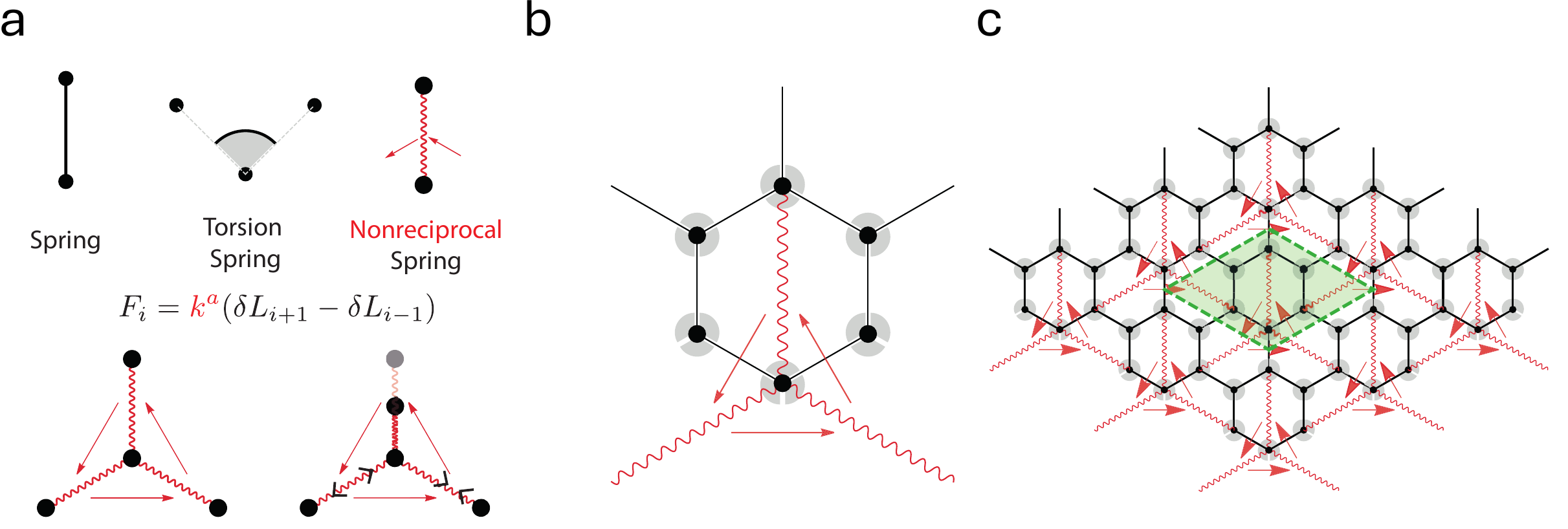}
\end{center}
\caption{ {\bf Theoretical representation of our 2D metamaterial}. {\bf (a)} We model our active lattice as a collection of longitudinal springs of stiffness $k$, torsional springs of stiffness $\kappa$, and non-reciprocal longitudinal bonds with activity $k^a$. Each non-reciprocal spring communicates with its neighbors with a sense shown by the thin red arrows. In the lower panel, we show an example of our non-reciprocal tension-length law. Compressing the top spring generates antisymmetric tensions in the neighboring bonds. {\bf (b,c)} A schematic of a single unit cell, and several cells, of our lattice.}

\label{COARSENING:TriHexSchematic}
\end{figure*}
Using the coarse-graining procedure described in Ref.~\cite{Binysh2025-xf} we can derive a relationship between odd modulus $K^o$ and the microscopic activity $k^a$ of the experimental lattice shown in Fig.~\ref{COARSENING:Fig1}. The unit cell of the lattice is modeled as a collection of longitudinal springs of stiffness $k$, torsional springs of stiffness $\kappa$, and non-reciprocal longitudinal bonds with activity $k^a$ (Fig.~\ref{COARSENING:TriHexSchematic}). We first construct the reciprocal space dynamical matrix $D({\bf q})$ for the ball-spring model and expand $D({\bf q})$ in the long wavelength limit, assuming internal degrees-of-freedom relax much faster that long-wavelength modes. Finally, from this reduced-order $D({\bf q})$, we read off 
the shear $\mu$, bulk $B$, and odd $K^o$ moduli 
in the relevant limit that $\kappa/k \ll 1 $ and $k^a/k \ll 1$, which are
\begin{align}
    K^o =\frac{3k^a}{4}, \\
    \mu = {2 \sqrt{3} \kappa}, \\
    B = \frac{k}{2 \sqrt{3}}.
\end{align}
In this limit, the dimensionless activity $\alpha = K^o/\mu =\frac{\sqrt{3} {k^a}}{8 \kappa } $ in units where $a=1$.
\subsection{Numerical simulations}\label{COARSENING:numericalsimulations}

Panels (c)-(d) of Fig.~\ref{COARSENING:Fig2} show the evolution of the inverse correlation length $\xi^{-1}$ and the spatially averaged shear magnitude $\langle|S|\rangle$, found by integration of the $\phi$ model of Eq.~\eqref{eq:phi_model} on a 1D grid of length L=1000 and discretization N=2000 and 2D grid with length $L_x=L_y=100$, discretization $N_x=200$, $N_y=200$, starting from random initial conditions using the py-pde package~\cite{py-pde}. Fig.~\ref{COARSENING:Fig3}e shows equivalent simulations for the 1D continuum model given by Eq.~\eqref{eq:beam} with $L=100$ and $N=1000$. 
In Fig.~\ref{COARSENING:Fig4}, we model the 2D metamaterial of Fig.~\ref{COARSENING:Fig1} as collection of masses and springs, in the arrangement described in Fig.~\ref{COARSENING:TriHexSchematic}. We model the rigid linkages using constraints implemented with Lagrange multipliers. The hinges are further described by bond bending viscoelastic forces and active forces. Using a velocity Verlet routine, we simulate a disk consisting of 5813 unit cells and in panels (d) and (e), we remove hexagons from the center to form an annular domain consisting of 5380 unit cells. 

\clearpage
\section{Supplementary text}
In this Supplementary Text, we derive the continuum equations Eq.~\eqref{eq:phi_model} and Eq.~\eqref{eq:beam} presented in the Main Text from continuum non-reciprocal elastodynamics (\S\ref{COARSENING:nonlinearoddelasticity}, \ref{COARSENING:fibre}). We next give primary and secondary stability analyses of Eq.~\eqref{eq:phi_model} 
to show that traveling nonlinear waves exist but are inevitably destabilized by secondary instabilities towards a cascade through mode space (\S\ref{COARSENING:PhiInstabilities}, \S\ref{COARSENING:CoarseningDynamics}). We then provide a simplified mean-field analysis of Eq.~\eqref{eq:phi_model} which captures this coarsening process via Eq.~\eqref{eq:NonlinearGrowthRateFirstOrder} of the Main Text (\S\ref{COARSENING:AmplitudeEquations}). We repeat this mean-field analysis for the beam model Eq.~\eqref{eq:beam} in \S\ref{COARSENING:fibre}. Finally, to capture the details of the edge modes shown in Fig.~\ref{COARSENING:Fig4} of the Main Text, we perform a detailed non-reciprocal Rayleigh wave calculation in \S\ref{COARSENING:Rayleigh} to show that edge localized elastic modes are unidirectionally amplified.

\section{Derivation of the $\phi$-model from nonlinear odd elasticity}\label{COARSENING:nonlinearoddelasticity}
In this section we derive a phenomenological model of nonlinear odd elasticity that captures wave coarsening in a minimal partial differential equation. We start by considering the continuum equations of an inertial, isotropic, viscoelastic odd solid, with density $\rho$, viscosity $\eta$, substrate friction $\gamma$, bulk and shear moduli $B$ and $\mu$, and odd moduli $K^o, A$~\cite{Scheibner_NatPhys2020}. The linear equations of odd elasticity in this context are given by
\begin{equation}
  B u_{k,ki}+\mu u_{i,kk}+\eta\dot{u}_{i,kk} -A u_{k,ki}\tau_{ij} + K^o \tau_{il}u_{l,kk}  = \gamma \dot{u}_i + \rho \ddot{u}_i,
  \label{eq:LinearDisplacement}
\end{equation}
where $u_i$ is the two-dimensional displacement field, $u_{i,k} := \partial_k u_i$, and $\tau$ is the Levi-Civita matrix. We use the simplest formulation of viscoelastic dissipation: a Kelvin-Voigt viscoelastic solid in which we only consider the shear viscoelasticity $\eta$~\cite{Landau}. We now seek to phenomenologically extend Eq.~\eqref{eq:LinearDisplacement} to capture nonlinear behaviour. Rather than turning to a fully nonlinear formulation of elasticity, we use a weakly nonlinear formalism in which we simply include nonlinearities that quench linear instabilities. This approach has been fruitful in describing soliton propagation in passive metamaterials~\cite{jinGuidedTransitionWaves2020}. We consider a passive stiffening nonlinearity in the shear, with strain energy 
\begin{equation}
  W(\epsilon_{ij}) = \frac{1}{2}B tr(\epsilon_{ij})^2 +\mu tr(S_{ij}^2)+\frac{\tilde{\mu}}{4} tr(S_{ij}^2)^2,
  \label{eq:StrainEnergy}
\end{equation}
where $\epsilon_{ij} = u_{i,j} + u_{j,i}$ is the linear elastic strain tensor, and $S$ denotes the traceless (pure shear) part of $\epsilon$, $S_{ij} = \epsilon_{ij} -\frac{1}{d} \epsilon_{kk}\delta_{ij}$, for spatial dimension $d=2$. To compute the modified elastic stresses, note that 
\begin{equation}
  \frac{\d S_{mn}}{\d \epsilon_{ij}} =\delta_{im} \delta_{jn} - \frac{1}{d} \delta_{mn}\delta_{ij}, 
\end{equation}
and so
\begin{equation}
  S_{mn} \frac{\d S_{mn}}{\d \epsilon_{ij}} = S_{ij}.
\end{equation}
With this identity, we have a modified elastic stress
\begin{equation}
  \begin{aligned}
    \frac{\d W}{\d \epsilon_{ij}}&= B\epsilon_{kk} + 2\mu S_{ij} + \tilde{\mu} S_{mn} S_{mn} S_{ij} \\
    &= B\epsilon_{kk} + (2\mu + \tilde{\mu} S_{mn} S_{mn}) S_{ij},
  \end{aligned}
\end{equation}
in other words, a rotationally invariant correction to the linear shear modulus. Using this elastic stress, we have a new equation of motion in $d=2$:
\begin{equation}
  B u_{k,ki}+\mu u_{i,kk}+\eta\dot{u}_{i,kk} -A u_{k,ki}\tau_{ij} + K^o \tau_{il}u_{l,kk} + \tilde{\mu}\partial_j \left( S_{mn} S_{mn} S_{ij}\right) = \gamma \dot{u}_i + \rho \ddot{u}_i.
  \label{eq:DisplacementPassive}
\end{equation}
Eq.~\eqref{eq:DisplacementPassive} represents a phenomenological equation for an isotropic odd elastic medium, with a dominant passive nonlinearity. We can derive a simplified version of Eq.~\eqref{eq:DisplacementPassive}, which captures the essential phenomenology observed in our experiments. Motivated by the observation that the bulk deformations of our metamaterial are small, we neglect the dilational term $u_{k,ki}$. Our elastic metamaterials also conserve angular momentum by construction, and so we additionally set $A=0$. We then have
\begin{equation}
\d^2_k \left( \mu u_i + \eta \dot{u}_i + K^{o} \tau_{il} u_l\right)+ \tilde{\mu} \d_k(S_{mn}S_{mn} S_{ik})= \gamma \dot{u}_i + \rho \ddot{u}_i.
\label{eq:DisplacementFormulation_dim}
\end{equation}
Writing Eq.~\eqref{eq:DisplacementFormulation_dim} in components:
\begin{align}
\d^2_k \left(\mu u_x +\eta \dot{u}_x + K^o u_y\right)+ \tilde{\mu}\d_k(S_{mn}S_{mn} S_{xk})= \gamma \dot{u}_x + \rho\ddot{u}_x, \label{eq:x} \\
\d^2_k \left(\mu u_y + \eta \dot{u}_y - K^o u_x\right)+ \tilde{\mu}\d_k(S_{mn}S_{mn} S_{yk})= \gamma \dot{u}_y+ \rho\ddot{u}_y. \label{eq:y}
\end{align}
Taking $\d_x \eqref{eq:x}-\d_y \eqref{eq:y} $, and $\d_y \eqref{eq:x}+\d_x\eqref{eq:y}$ to express these equations in terms of the two shear components $S_1=\frac{1}{2}(\d_x u_x -\d_y u_y)$, $S_2 = \frac{1}{2}(\d_x u_y + \d_y u_x)$, we obtain a single equation in the complex variable $\phi=S_1 + i S_2$:
\begin{equation}
\nabla^2 \left( (\mu-iK^o)\phi+ \eta\dot{\phi} + \tilde{\mu}|\phi|^2 \phi \right)= \rho \ddot{\phi} + \gamma \dot{\phi}.
\label{eq:phi_model_dimensional}
\end{equation}
We now non-dimensionalize Eq.~\eqref{eq:phi_model_dimensional} using viscous length and time scales and a redefinition of the shear strain amplitude:
\begin{equation}
\begin{aligned}
\mathrm{Viscous\ Timescale:}\  T_{v} = \frac{\eta}{\mu}, \quad 
\mathrm{Viscous\ Lengthscale:} \ X_{v}= \sqrt{\frac{\mu}{\rho}} \frac{\eta}{\mu}, \quad
\mathrm{Amplitude\ Rescaling:} \  \phi \rightarrow \sqrt{\frac{\tilde{\mu}}{\mu}}\phi,
\end{aligned}
\end{equation}
resulting in
\begin{align}
\nabla^2 \left[ (1-i\alpha)\phi+ \dot{\phi} + |\phi|^2 \phi \right]= \ddot{\phi} + {\Gamma} \dot{\phi},
\label{eq:phi_modelSI}
\end{align}
where $\alpha:= K^0/\mu$ is the activity and $\Gamma = \gamma \eta/(\rho \mu)$ is the dimensionless friction. We dub Eq.~\eqref{eq:phi_modelSI} the $\phi$ model--- in the frictionless case of $\Gamma=0$ it is Eq.~\eqref{eq:phi_model} of the Main Text---and we now analyze its dynamics and stability. 

\section{Travelling patterns and secondary instabilities in the $\phi$ model}\label{COARSENING:PhiInstabilities}
In this section we analyze the linear and nonlinear dynamics of the $\phi$-model Eq.~\eqref{eq:phi_modelSI}.
\subsection{Linear stability analysis}
\label{sec:Analysing}
\begin{figure*}[t!]
\begin{center}
\includegraphics[width=1\linewidth,trim=0cm 0cm 0cm 0cm]{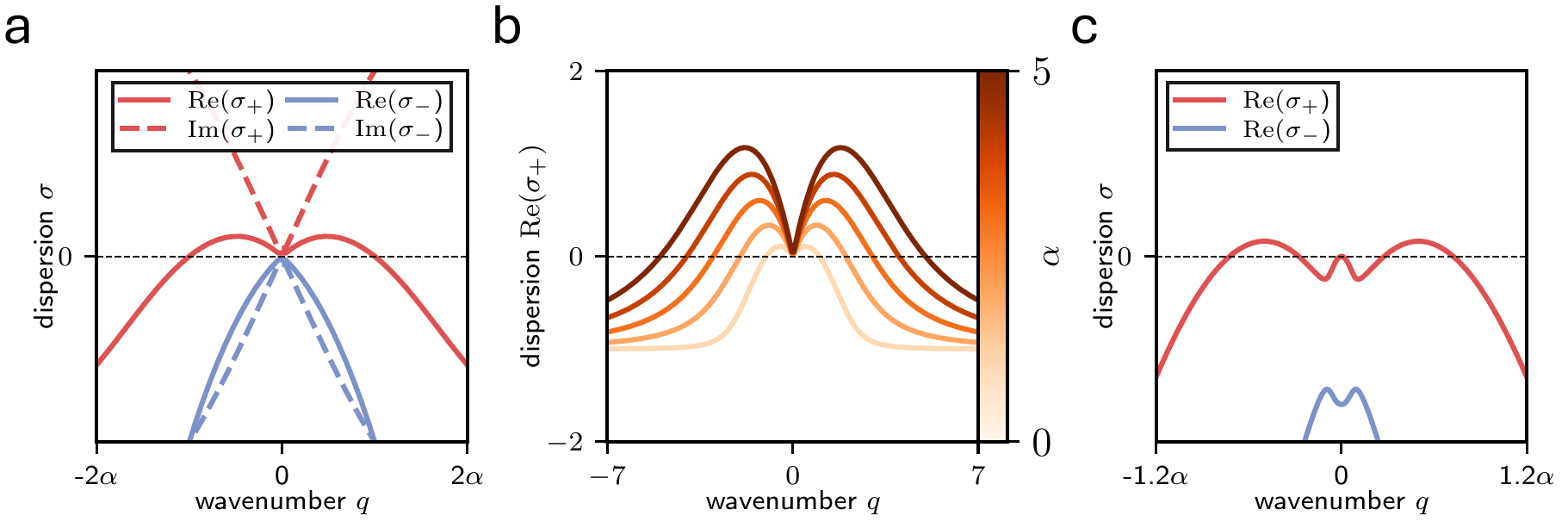}
\end{center}
\caption{\linespread{1.1}\selectfont{}
\textbf{Features of the $\phi$ model dispersion.} {\bf (a)} Real and imaginary parts of the dispersion when $\Gamma=0$. One mode is damped (blue), but we also find a band of unstable wavenumbers from $q=0$ to $q=\pm \alpha$ (red). The imaginary part of the dispersion is explicitly odd in $q$, breaking left-right symmetry. {\bf (b)} Several dispersions for varying $\alpha$, with $\Gamma=0$. {\bf (c)} Including friction $\Gamma$ gaps the spectrum at $q=0$. Here $\alpha=1$, $\Gamma=0.2$.}
\label{fig:PhiDispersion}
\end{figure*}
Linearizing Eq.~\eqref{eq:phi_modelSI} and taking $\phi = A \exp(i qx + \sigma t)$ we have the dispersion
\begin{align}
\sigma^2 + \Gamma \sigma +q^2(1- i \alpha + \sigma) &= 0 \label{eq:phimodelLinear},
\end{align}
with solution 
\begin{align}
\sigma_{\pm}= 
\frac{1}{2} \left(-q^{2}-\Gamma \pm 
\sqrt{ -4 q^{2} + q^{4} + 4 iq^{2} \alpha + 2 q^{2} \Gamma + \Gamma^{2} }
\right).
\label{eq:Dispersion}
\end{align}
This dispersion is shown in Fig.~\ref{fig:PhiDispersion}. We find two branches of roots, $\sigma_\pm(q)$. The branch $\sigma_-(q)$ is always damped. However, $\sigma_+(q)$ posseses a band of unstable wavenumbers. The roots of $\mathrm{Re}[\sigma_+(q)]$ can be found by inspection as either $q^*=0$, or as the wavenumbers $q^*$ for which
\begin{equation}
\begin{aligned}
   {q^*}^2 - \alpha q^* + \Gamma = 0,\\ 
   q^*= \frac{1}{2} \left( \alpha \pm \sqrt{ {\alpha}^2- 4 \Gamma}\right ).
   \label{eq:roots}
\end{aligned}
\end{equation}
The threshold of instability is then given by the condition that these roots are degenerate, $\alpha = 2 \sqrt{\Gamma}$. Focusing first on the case of no friction, $\Gamma=0$, we find $\mathrm{Re}[\sigma(q)]$ has two zeros at $q=0, \alpha$, and a band of unstable wavenumbers growing continuously from the origin. For small $\alpha$, the dispersion is approximately $\sigma \approx \frac{1}{2}q(\alpha-q)+ iq$: an oscillatory long-wave, or type-IIo, instability~[Fig.~\ref{fig:PhiDispersion}(a,b)]~\cite{cross_pattern_1993}, with the threshold for instability at $\alpha=0$. Including friction $\Gamma$, we additionally introduce a new root in $\mathrm{Re}(\sigma_+)$, gapping the spectrum at $q=0$ [Fig.~\ref{fig:PhiDispersion}(c)]. In the limit of small $\Gamma$, where we assume $q \sim \Gamma \ll \alpha$, the leading order behaviour of this root is given by $q^* \sim \Gamma/\alpha$.

Type-II instabilities drive coarsening dynamics across a wealth of systems: in equilibrium~\cite{Bray2002-tn} and non-equilibrium phase separation~\cite{Saha_PRX2020,Fruchart2021non,suchanek_entropy_2023,braunsNonreciprocalPatternFormation2024, sahaEffervescenceBinaryMixture2025}, sedimentation phenomena~\cite{ramaswamyCoarsening}, active~\cite{chatterjeeInertiaDrivesFlocking2021} and driven~\cite{nepomnyashchy_coarsening_2015,shklyaevLongwaveInstabilitiesPatterns2017} fluids, and crystal growth~\cite{politi_when_2004}. The canonical type-II dispersion is stationary (no oscillations), and respects the reflection symmetry $q\rightarrow-q$: $\sigma(q)\sim q^2 - q^4$~\cite{cross_pattern_1993}. By contrast, non-reciprocal elastic couplings explicitly break this symmetry~\cite{Scheibner_NatPhys2020}, allowing for a  dispersion that is linear in $q$, and the inertial character of our traveling waves gives the instability an oscillatory component.

\subsection{Nonlinear travelling waves}
Beyond linear instability, we find nonlinear travelling wave states that solve Eq.~\eqref{eq:phi_modelSI}. By inspection we find
\begin{equation}
\begin{aligned}
\phi(x,t) &= A(q) \exp\left[ i\left( \omega(q) t \pm q x\right) \right], \\
\omega(q) &= \frac{q^2 \alpha}{\Gamma + q^2}, \\
A(q) &=\sqrt{\left(\frac{\omega(q)}{q} \right)^2-1} =  \sqrt{ \left( \frac{ \alpha q}{\Gamma + q^2}\right)^2-1}.
\label{eq:NESS}
\end{aligned}
\end{equation}
We have a family of travelling waves of amplitude $A(q)$. The maximal amplitude occurs at $q^* = \sqrt{\Gamma}$, and as the friction $\Gamma \rightarrow 0 $, this amplitude $q^* \rightarrow 0$: ever-increasing wave amplitudes exist at ever-larger length scales, until eventually saturating at the system size. Note that the amplitude of these states matches the criteria for linear instability. Once a mode has gone linearly unstable, a nonlinear wave state immediately exists, with an amplitude that grows continuously from zero, as the square root of a bifurcation parameter $(\omega/q)^2-1$. This behaviour is typical of a supercritical Hopf bifurcation. 

\subsection{Secondary stability analysis}

\begin{figure*}[t!]
\begin{center}
\includegraphics[width=0.6\linewidth,trim=0cm 0cm 0cm 0cm]{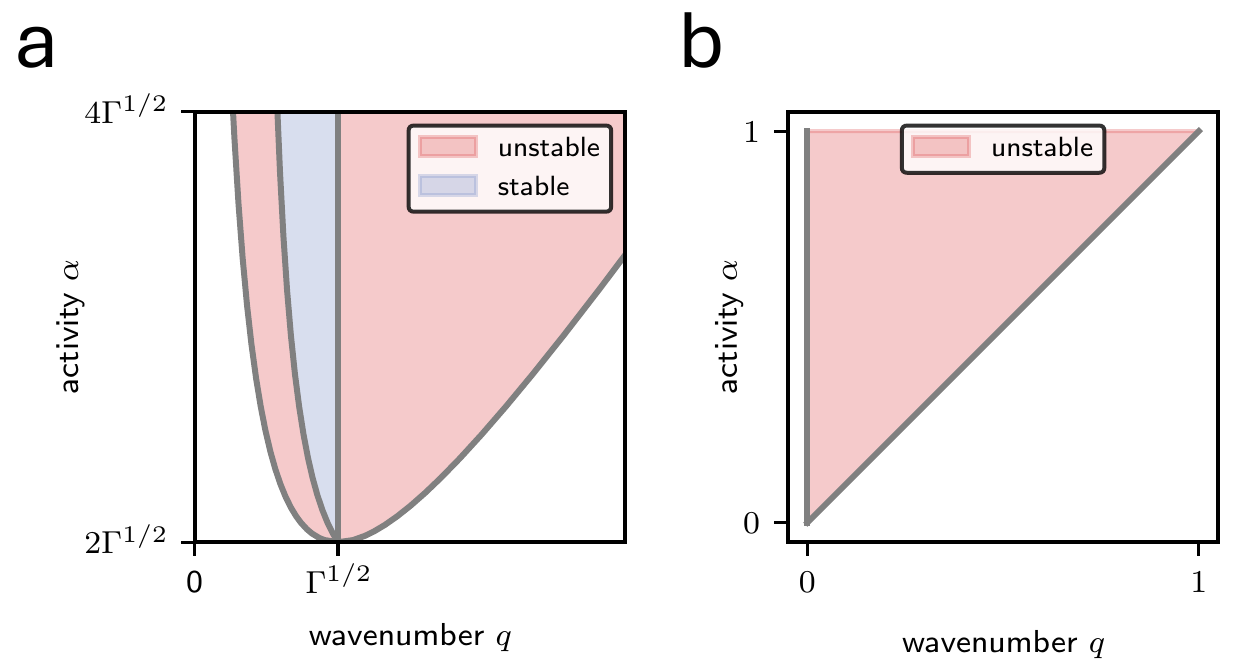}
\end{center}
\caption{\linespread{1.1}\selectfont{}
\textbf{Secondary Stability of the $\phi$ model}. Stability of the nonlinear traveling patterns hosted by Eq.~\eqref{eq:phi_modelSI}. Outer gray curves indicate the wavenumbers $q$ which becomes linearly unstable as $\alpha$ is increased. Inner gray curves show the boundaries of additional secondary instabilities. {\bf (a)} With friction $\Gamma$. {\bf (b)} Without friction, $\Gamma=0$.}
\label{fig:SecondaryStability}
\end{figure*}

We have a family of wave solutions given by Eq.~\eqref{eq:NESS}, and we now perform a secondary stability analysis about these wave states~\cite{cross_pattern_1993}.
We consider a perturbed wave
\begin{equation}
\phi(x,t) =\phi_q(x,t) +\delta\phi(x,t) := A(q) e^{ i \left(\omega t - q x\right)} + \delta\phi(x,t)
\label{eq:ansatz}
\end{equation}
where we define a travelling wave base state $\phi_q(x,t)$ from Eq.~\eqref{eq:NESS}, indexed by base wavenumber $q$, and a perturbation about this state $\delta\phi(x,t)$. Substituting this ansatz into Eq.~\eqref{eq:phi_modelSI} our linearized dynamics reads
\begin{equation}
\delta \ddot{\phi} +\Gamma \dot{\phi} = \nabla^2 \left( (1- i \alpha) \delta \phi + \delta \dot{\phi} + 2 |\phi_q|^2 \delta \phi + \phi_q^2 \delta \phi^* \right).
\label{eq:SecondaryFriction}
\end{equation}
 The term $\phi_q^2 \delta \phi^* = A(q)^2 \exp{2i (\omega t - qx)}$ will mix modes. 
 We make two observations. First, it is natural to assume that the perturbation rotates with the same frequency as the base state, $\delta\phi \sim e^{i\omega t}$. Second, Bloch's theorem tells us we can label spatial perturbations by a second wavenumber $k$, which we measure relative to the base wavenumber $q$ as $q\pm k$~\cite{cross_pattern_1993}. The wavenumber $k$ is a perturbation about $q$: when $k =0$ we are perturbing with the same wavenumber $q$ as the base state. In summary, we try the ansatz 
\begin{align}
\delta \phi = e^{i(\omega t-q x)} \left(a_+ (t) e^{i k x} + a^*_{-} (t) e^{-i k x} \right) = e^{i\omega t} \left(a_+ (t) e^{i (k - q) x} + a^*_{-} (t) e^{-i (k+q) x} \right),  \\
\delta \phi^* = e^{-i(\omega t-q x)} \left(a^*_+ (t) e^{-i k x} + a_{-} (t) e^{i k x} \right) = e^{-i\omega t} \left(a^*_+ (t) e^{-i(k - q) x} + a_{-}(t) e^{i(k+q) x} \right), 
\label{eq:Secondary}
\end{align}
and substitute into Eq.~\eqref{eq:SecondaryFriction}. Gathering exponentials we have
\begin{equation}
\begin{aligned}
\delta \ddot{\phi} +\Gamma \dot{\phi} = 
\left[ -\omega^2 a_{+} +  2 i \omega \dot{a}_{+}  + \ddot{a}_{+} + i \Gamma \omega a_{+} + \Gamma \dot{a}_{+} \right] e^{i\omega t} e^{i (k - q) x}  \\
+ \left[ -\omega^2 a^*_{-} +  2 i \omega \dot{a}^*_{-}  + \ddot{a}^*_{-} + i \Gamma \omega a^*_{-} + \Gamma \dot{a}^*_{-} \right] e^{i\omega t}e^{-i (k + q) x}  \\
\end{aligned}
\end{equation}
and
\begin{equation}
\begin{aligned}
\nabla^2 \left[ (1- i \alpha) \delta \phi + \delta \dot{\phi} + 2 |\phi_q|^2 \delta \phi + \phi_q^2 \delta \phi^* \right]=
\nabla^2 [ 
 \left( (1- i \alpha) a_{+} +  i \omega a_{+} + \dot{a}_{+} + 2A^2a_{+} + A^2 a_{-}  \right) e^{i\omega t} e^{i (k - q) x}  \\
+\left( (1- i \alpha) a^*_{-} +  i \omega a^*_{-} + \dot{a}^*_{-} + 2A^2a^*_{-} + A^2 a^*_{+}  \right) e^{i\omega t} e^{-i (k + q) x} ],
\end{aligned}
\end{equation}
from which we find
\begin{equation}
\begin{aligned}
(i \Gamma \omega-\omega^2 )a_+ + (\Gamma + 2i \omega ) \dot{a}_+ + \ddot{a}_+ = - (\delta k-q)^2 \left[ \left(1+ i (\omega - \alpha) +  2 A(q)^2\right) a_+ + \dot{a}_+ + A(q)^2 a_-\right], \\
(-i \Gamma \omega-\omega^2 ) a_{-} + (\Gamma - 2i\omega) \dot{a}_- + \ddot{a}_- = - (\delta k+q)^2 \left[ \left(1- i(\omega- \alpha) +2 A(q)^2\right) a_- + \dot{a}_- + A(q)^2 a_+\right].
\end{aligned}
\end{equation}
Taking $a_+ = \tilde{a}_+ e^{\sigma t},a_{-}=\tilde{a}_- e^{\sigma t}$ we have the matrix equation
\begin{equation}
\left(
\begin{smallmatrix}
 (k-q)^2 \left(i (\omega -\alpha )+2 A^2+1\right)+\sigma  (\Gamma +2 i \omega )+i \omega  (\Gamma +i \omega )+\sigma  (k-q)^2+\sigma ^2 & A^2 (k-q)^2 \\
 A^2 (k+q)^2 & (k+q)^2 \left(i \alpha +2 A^2-i \omega +1\right)+\sigma  (\Gamma -2 i \omega )+\omega  (-\omega -i \Gamma )+\sigma  (k+q)^2+\sigma ^2 \\
\end{smallmatrix}
\right)
\left(
\begin{smallmatrix} {\tilde{a}_+} \\ {\tilde{a}_-}
\end{smallmatrix}\right) \\ =0,
\label{eq:Secondary3}
\end{equation}
whose determinant gives the growth rates $\sigma$ of perturbations $k$ to a travelling wave at wavenumber $q$ (Fig.~\ref{fig:SecondaryStability}). For finite friction $\Gamma$ we find a band of modes about the dominant wavenumber $q=\Gamma^{1/2}$ of width $\sim \Gamma^{1/2}$ which are stable. Waves outside this band are vulnerable to a Benjamin-Feir instability. As $\Gamma \rightarrow 0$ this stable band vanishes, and at zero friction we find that all modes are unstable towards phase slips.

\section{Observation of coarsening dynamics in the $\phi$ model}\label{COARSENING:CoarseningDynamics}
\begin{figure*}[t]
\centering
\includegraphics[width=1\columnwidth]{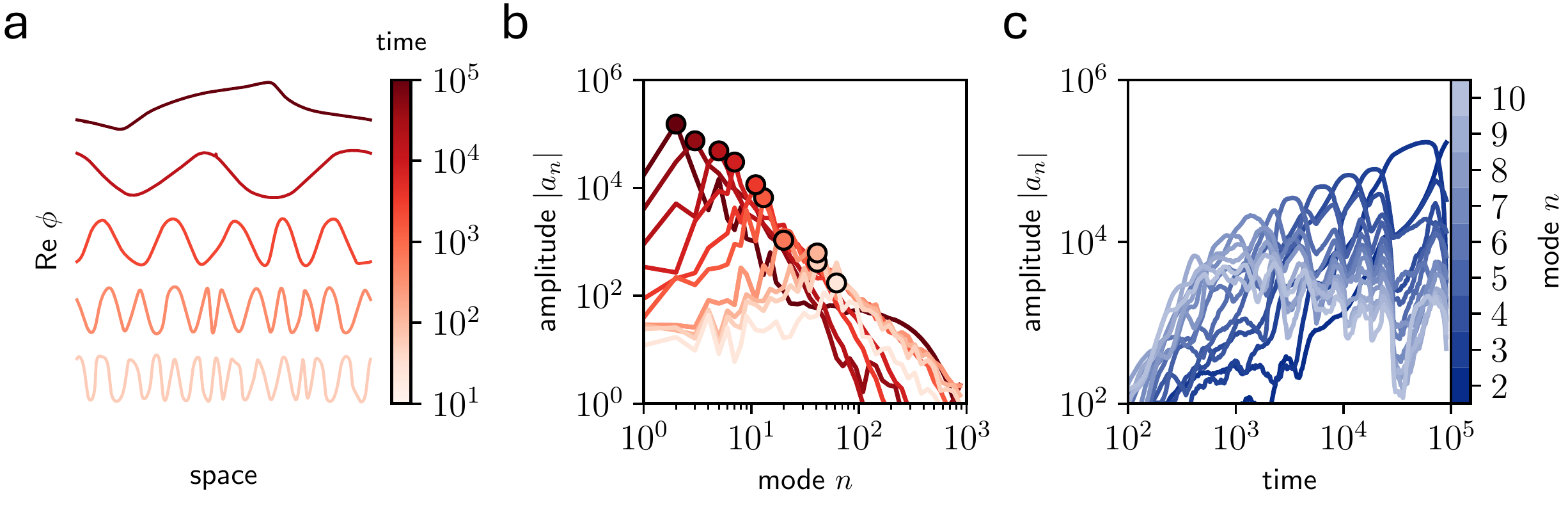}
\caption{\textbf{Coarsening in the $\phi$ model.} A numerical simulation of Eq.~\eqref{eq:phi_model} for $\alpha=1$, boxsize $L=1000$ and space is discretized in $N=2000$ steps.{\bf (a)} The real part of $\phi$ at five orders of time. {\bf (b)} The amplitude spectrum at different times, with a marker indicated the peak of the spectrum. {\bf (c)} The ten lowest mode amplitudes over time.}
\label{fig:Coarsening1}
\end{figure*}
The above stability analyses suggest a two-step process. First, non-reciprocity will pick a travelling pattern of given wavenumber through a type-II linear instability. Next, nonlinear couplings will destabilize this mode. What is the eventual fate of this travelling pattern?
In Fig.~\ref{fig:Coarsening1}(a), we show simulation snapshots of the $\phi$-model in a 1D geometry under periodic boundaries. Above the threshold for linear instability, we initially see a dominant wavelength emerge. Over time, however, this mode undergoes a sequence of secondary instabilities, which cascade towards ever lower mode number, and ever larger amplitude. This observation is confirmed by the amplitude spectrum, whose peak shifts to lower modes and larger amplitudes over time~[Fig.~\ref{fig:Coarsening1}(b,c)]. In one dimension, the cascade of wavenumbers is mediated by phase slips, events where the dominant wavenumber suddenly decays and a lower mode takes over [Fig.~\ref{fig:Coarsening1}(c)]. Tracking the inverse correlation length of such data for varying $\alpha$ in both 1D and 2D reveals that this coarsening process occurs via a power law, which we show in Fig.~\ref{COARSENING:Fig2} of the Main Text.

\section{Mean field approximation to the $\phi$ model}\label{COARSENING:neglectingterms}
In the Main Text, we approximated the full mode equation corresponding to a nonlinear odd elastic solid by neglecting cross coupling between modes, and reducing from a second order dynamics to a first-order dynamics that captures the coarsening process but neglects fast oscillations. Here we provide numerical results that justify the neglect of cross couplings, effectively producing a mean field model. Next, we apply a multiple scales approach to reduce this second order mean field model to a first order model.

\subsection{Reduction to the mean field model}
 We write Eq.~\eqref{eq:phi_modelSI} in mode space, $\phi = \sum_n a_n e^{iq_nx}$, and decompose the nonlinear term into an onsite nonlinearity, a bath interaction term and a phase dependent term:
\begin{figure}[h]
\centering
\includegraphics[width=0.7\columnwidth]{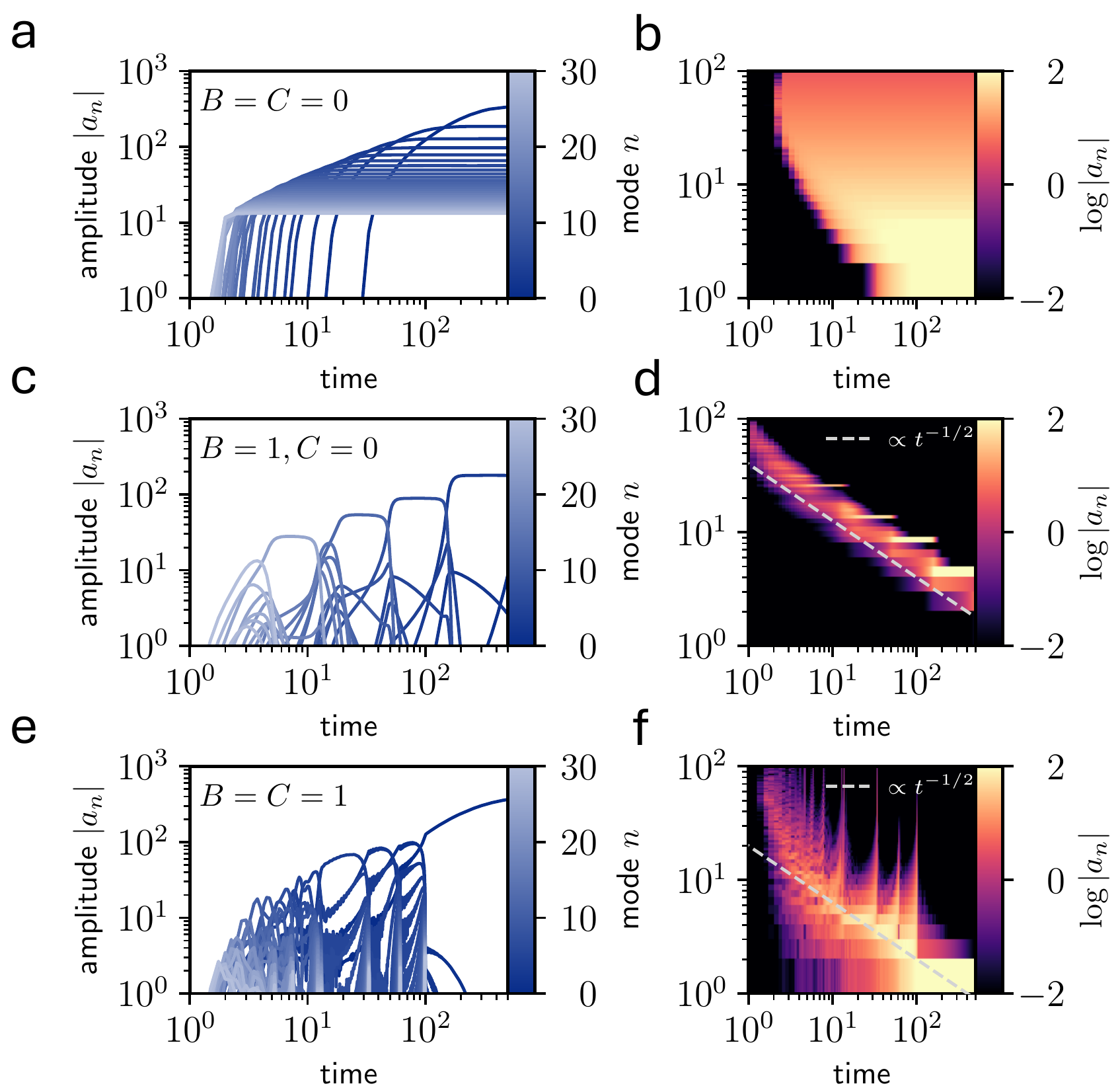}
\caption{
\textbf{ Comparison between full and approximated mode equation.} Simulations of Eq.~\eqref{COARSENING:separatenonlinearterms} with {\bf (a,b)} $B=C=0$, {\bf (c,d)} $B=1,C=0$ and {\bf (e,f)} $B=C=1$, using $N=100$ and $L=\alpha=50$. The kymograph representation of the same data in {\bf (b,d,f)} shows the mode distribution over time and a power-law function $\propto t^{-1/2}$. 
}
\label{COARSENING:fig_neglecting}
\end{figure}
\begin{align}
\ddot{a}_n(t)&=-q_n^2 \left[ (1-i\alpha)a_n + \dot{a}_n +
\underbrace{|a_n|^2a_n}_\text{onsite}+
2B\underbrace{\sum_{m\neq n} |a_m|^2 a_n}_\text{bath}+
C\underbrace{\sum_{i\neq j \neq n} a_i a^{*}_j a_{n-i+j}}_\text{phase dependent}
\right],
\label{COARSENING:separatenonlinearterms}
\end{align}

\noindent where we have parametrized the bath and phase dependent terms by $B,C\in \{0,1\}$.
In order to assess the validity of the approximation in the main text that $C=0$, we integrate this equation using a spectral method for the cases that $B=C=0$, $B=1,C=0$, and $B=C=1$. In the case that $B=C=0$, the equation represents a collection of uncoupled nonlinear oscillators (Fig.~\ref{COARSENING:fig_neglecting}a). Although the lowest mode eventually dominates, the absence of interaction terms means that there is a broad distribution of modes at all times (Fig.~\ref{COARSENING:fig_neglecting}b).

When $B=1,C=0$, we recover the approximated model discussed in the main text which captures the features of the experimental data observed in Figs~\ref{COARSENING:Fig1} and~\ref{COARSENING:Fig4} (Fig.~\ref{COARSENING:fig_neglecting}c). As one mode starts to dominate, the next one grows and eventually takes over until only the lowest mode is left.
As demonstrated in the next section, the existence of a nonlinear growth rate implies that only a narrowly distributed band of modes is excited at any time. During coarsening, this band 
shifts at a rate proportional to $\propto t^{-1/2}$(Fig.~\ref{COARSENING:fig_neglecting}d).

Finally, when $B=C=1$ we recover the full model where modes resonate during the mode evolution (Fig.~\ref{COARSENING:fig_neglecting}e). Nonetheless, the dominant mode still coarsens over time and tends to the same $-1/2$ power-law as the mean-field model (Fig.~\ref{COARSENING:fig_neglecting}f).

\subsection{Reduction from second to first order}\label{COARSENING:AmplitudeEquations}
Having established that the wave coarsening can quantitatively be captured in the mean field setting, we analyze this simplified model. For ease of notation, in this section we use $q=q_n = 2 \pi n/L$ interchangeably where it is clear from context. Assuming that the on-diagonal bath term dominates couplings between modes in Eq.~\eqref{COARSENING:separatenonlinearterms} yields 
\begin{align}
\ddot{a}_n(t)&=  -q_n^2 \left[ (1-i\alpha)a_n + \dot{a}_n + \left( 2 \sum_{m} |a_m|^2\right)a_n \right].
\label{eq:k-space-meanfield} 
\end{align}
Eq.~\eqref{eq:k-space-meanfield} is second order in time. However, we now show that a multiple scales approach can reduce the dynamics to an effectively first order description. We make an exponential coordinate transformation, $a_n(t) = e^{\sigma_{n}(t)}$. The growth rate of amplitudes $\dot{|a_n|}$ and instantaneous frequencies $\omega_n = \dot{\theta}_n$ of $a_n= |a_n|e^{i\theta}$ are recovered via the relation $\sigma_n = \dot{|a_n|}/|a_n| + i \omega_n$. In terms of this transform, Eq.~\eqref{eq:k-space-meanfield} reads
\begin{equation}
    \dot{\sigma}_{n}^{2} + \ddot{\sigma}_{n} = -q_n^{2} \left( (1- i \alpha) + \dot{\sigma}_{n} + 2 \sum_{m} |a_{m}|^{2} \right).
\label{eq:exponential_transform_full}
\end{equation}
Here, $\dot{\sigma}_{n}$ is a time dependent function with instantaneous frequency $\omega_{n} := \text{Im}(\dot{\sigma}_{n})$ and instantaneous exponential growth $g_{n} := \text{Re}(\dot{\sigma}_{n})$. We note that there are two timescales in Eq.~\eqref{eq:exponential_transform_full}. In the absence of the $\ddot{\sigma}_n$ term, Eq.~\eqref{eq:exponential_transform_full} is an algebraic relation for $\dot{\sigma}_n$. We hypothesize that this algebraic relation gives the dynamics of $\dot{\sigma}_n$ on a slow timescale. We therefore split $\dot{\sigma}_n$ as
\begin{align}
     \dot{\sigma}_{n} = \dot{\langle \sigma \rangle}_{n} + \dot{\delta}_{n}
     \label{eq:Split}
\end{align}
where we define $\dot{\langle \sigma \rangle}_{n}$ as a slow time variable which satisfies
\begin{align}
    \dot{ \langle \sigma \rangle }_{n}^{2} =& -q_n^{2} \left( (1- i \alpha) + \dot{ \langle \sigma \rangle}_{n} + 2 \sum_{m} |a_{m}|^{2} \right)
    \label{eq:SlowTimeSolution}
\end{align}
with a perturbative remainder $\dot{\delta}_{n}$ capturing the fast time dynamics. We look at the stability of small perturbations, such that we assume $\dot{\delta}_{k}$ is small. Substituting Eq.~\eqref{eq:Split} into Eq.~\eqref{eq:exponential_transform_full} and using Eq.~\eqref{eq:SlowTimeSolution} yields 
\begin{align}
     \ddot{\langle \sigma \rangle}_n + \ddot{\delta_n} + 2\dot{\langle \sigma \rangle}_n\dot{\delta_n} + \dot{\delta_n}^{2} =0.
\end{align} 
We have that $\dot{\delta}^2$ is small, and by assumption of multiple timescales $\ddot{\langle \sigma \rangle}$ is also small. We obtain
\begin{align}
    \ddot{\delta}_{n} + 2 \dot{\langle \sigma \rangle}_{n} \dot{\delta}_{n}  &= 0.
\label{eq:ODE_delta} 
\end{align}
Equation~\eqref{eq:ODE_delta} is now solvable on fast timescales. In summary, we obtain 
\begin{align}
    \dot{ \langle \sigma \rangle}_{n} &= -\frac{q_n^{2}}{2} + \frac{q_n}{2} \sqrt{ q_n^{2} - 4 \left( (1-i \alpha) + 2 \sum_{m} |a_{m}|^{2} \right) }, \\
    \dot{\delta}_{n} &= \left( \dot{\delta}_{n}(t_{0}) e^{2 \langle \sigma \rangle_{n}(t_{0})} \right) e^{- 2 \langle \sigma \rangle_{n} }. 
    \label{eq:dotsigma} 
\end{align} 
Taylor expanding from time $t_{0}$ to obtain $\dot{\delta}_{n}(t_{0}+\Delta t) = \dot{\delta}_{n}(t_{0}) e^{ - 2 \dot{ \langle \sigma \rangle }_{n}(t_{0}) \Delta t }$. Since $g_{n} \approx \text{Re}(\dot{\langle \sigma \rangle}_{n})$ gives the instantaneous exponential growth and $\omega_{n} \approx \text{Im}(\dot{\langle \sigma \rangle}_{n})$, we have $\dot{\delta}_{n}(t_{0} + \Delta t) = \dot{\delta}_{n}(t_{0}) e^{- 2 g_{n}(t_{0}) \Delta t - 2 i \omega_{n}(t_{0}) \Delta t} $. Therefore, for a growing mode, the perturbation decays exponentially fast in an oscillatory manner. The hypothesis encoded in Eq.~\eqref{eq:Split} is self-consistent: after a short time interval we find that $\dot{\delta} = 0$, and in the coarsening regime $\dot{\sigma}_{n} \rightarrow \dot{\langle \sigma \rangle}_{n}$.

We now consider the limit of large nonlinearity $\sum_{m} |a_{m}|^{2}$. Our linear stability analysis Eq.~\eqref{eq:roots} shows that amplified modes are bounded by $\alpha$. We therefore Taylor expand the root in Eq.~\eqref{eq:dotsigma} by taking $1+2 \sum_{m} |a_{m}|^{2}$ to be large compared to $\alpha$ and $q_n$. We obtain 
\begin{align}
    \dot{\sigma}_{n} \approx - \frac{q_n^{2}}{2} + i{q_n}  \sqrt{ 1+2 \sum_{m} |a_{m}|^{2}} + \frac{  \alpha q_n }{ 2  \sqrt{1 + 2 \sum_{m} |a_{m}|^{2}  } }.
\end{align}
\begin{figure*}[t!]
\begin{center}
\includegraphics[width=0.35\linewidth,trim=0cm 0cm 0cm 0cm]{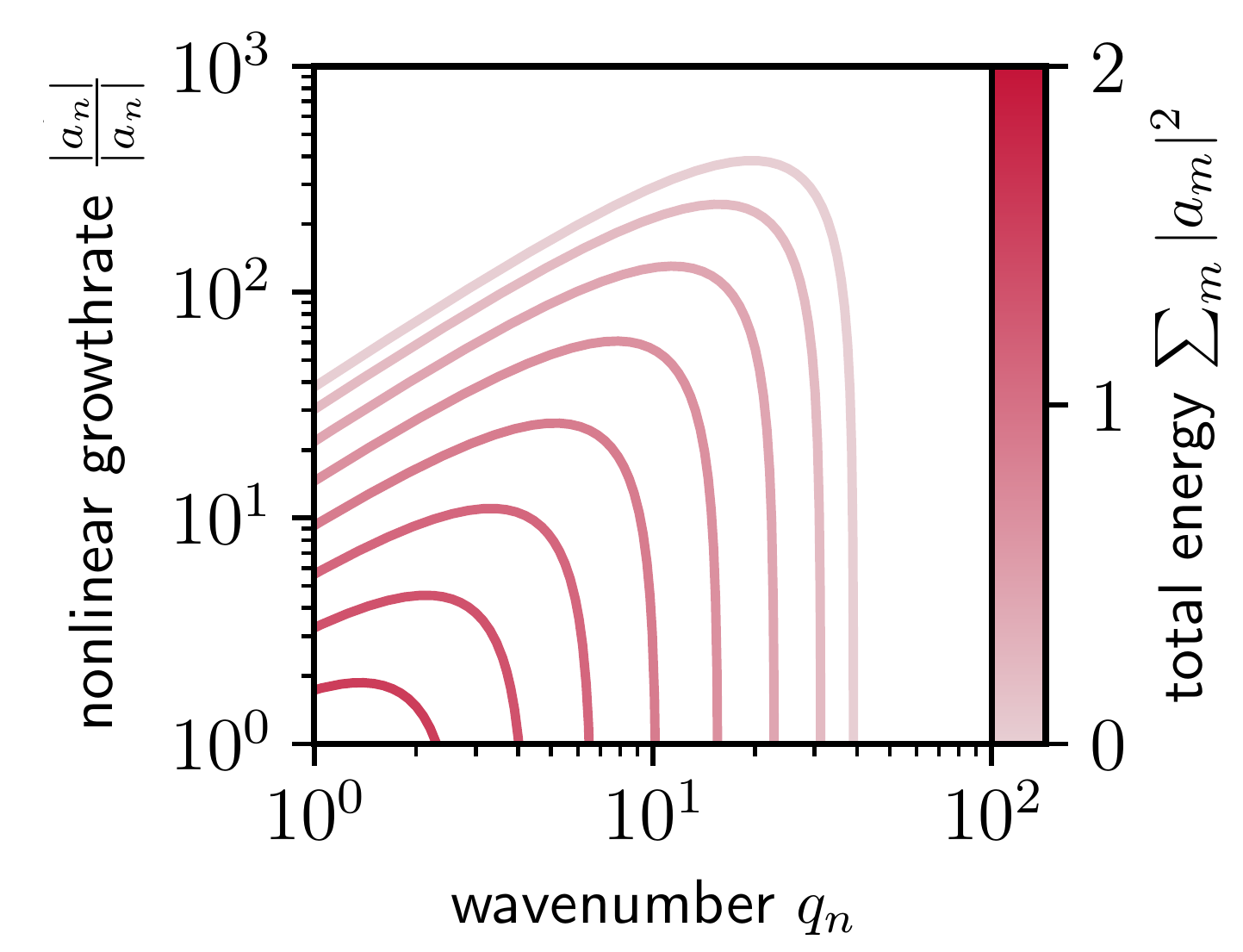}
\end{center}
\caption{{\textbf{The nonlinear growth rate.}} As the total elastic energy $1+2 \sum_{m} |a_{m}|^{2}$ changes over time, the nonlinear growth rate $\frac{|\dot{a_n}|}{|a_n|}$ decreases and shifts to lower wavenumbers.}
\label{fig:NonlinearGrowthrate}
\end{figure*}
We take the real part of $\langle\dot{\sigma}\rangle_{n}$ to obtain the instantaneous exponential growth rate and the imaginary part to obtain the instantaneous dispersion. We obtain for the frequency
\begin{equation}
    \omega_n(t) = q_n \sqrt{1 + 2 \sum_{m} |a_m(t)|^2},
    \label{eq:NonlinearFrequency}
\end{equation}
and for the amplitude $|a_n|$
\begin{equation}
    \dot{|a_n|} = \frac{1}{2} \left( \frac{ \alpha }{ \sqrt{1+ 2 \sum_{m} |a_m|^2} }q_n - {q_n^{2}} \right)|a_n|,
\label{eq:FirstOrder}
\end{equation}
where Eq.~\eqref{eq:FirstOrder} corresponds to Eq.~\eqref{eq:NonlinearGrowthRateFirstOrder} of the Main Text for the case where $p =1$, and represents a first order dynamics for the nonlinear growth rate of mode amplitude $|a_n|$. The structure of Eq.~\eqref{eq:FirstOrder} is identical to the linear dispersion Eq.~\eqref{eq:Dispersion}, except for a dynamic rescaling of $\alpha$ by a factor $\sqrt{1+ 2 \Sigma_m |a_m|^2}$. Intuitively, as this bath term grows, the effective maximal growth rate of the linear dispersion is shifted towards $q
\rightarrow 0$, and ever-lower modes are preferentially amplified at ever lower rates (Fig.~\ref{fig:NonlinearGrowthrate}). To solve Eq.~\eqref{eq:FirstOrder}, we multiply by $|a_{n}|$ and obtain
\begin{equation}
    \frac{d}{dt}(|a_n|^2) = \left( \frac{ \alpha }{ \sqrt{ 1+ 2 \sum_{m} |a_m|^2} }q_n - {q_n^{2}} \right)|a_n|^2.
    \label{eq:amplitude_square_form}
\end{equation}
Appealingly Eq.~\eqref{eq:amplitude_square_form} suggests an argument for the coarsening exponents. Suppose coarsening occurs via the motion of a tightly peaked packet of wavenumbers, such that the mean wavenumber is much larger than the packet variance. We then convert Eq.~\eqref{eq:amplitude_square_form} into an equation involving the total amplitude and central wavenumber $Q$ only. We proceed by obtaining the first two moments of the normalized distribution $2 |a_{n}|^{2}/A$ as
\begin{align}
    2 \sum_{m} |a_{m}|^{2} &:= A, \\
    2\sum_{m} q_m |a_{m}|^{2} &= A \sum_{m} q_m \frac{2 |a_{m}|^{2}}{A} = A Q, \\
    2\sum_{m} q_m^{2} |a_{m}|^{2} &= A \sum_{m} q_m^{2} \frac{2 |a_{m}|^{2}}{A} = A ( Q^{2} + s^{2} ). 
\end{align}
Given the concentrated mode distribution, we have that its variance $s$ is much smaller than the square of the central wavenumber $Q$, such that $\langle q^{2} \rangle = Q^{2} + s^{2} \approx Q^{2}$. Using this, we sum both sides of Eq.~\eqref{eq:amplitude_square_form} over all wavenumbers $q$ and multiply by $2$ to obtain
\begin{equation}
\dot{A} = \alpha Q(t) A^{1/2}(t) -Q(t)^2 A(t). 
\end{equation} 
Finally we make the ansatz $A(t) \sim t^{\chi}$, $Q(t) \sim t^\beta$. We find
\begin{equation}
    t^{\chi-1} \sim t^{2\beta + \chi} + t^{\beta + \frac{1}{2}\chi},
\end{equation}
which only has a solution if the powers are the same for each term. From the equation $\chi - 1 = 2 \beta + \chi$ we obtain $\beta = -1/2$. From the equation $\chi - 1 = \beta + \frac{1}{2} \chi $ we obtain $\chi = 1$. We therefore predict a characteristic scale for the peak wavenumber, and amplitude, of
\begin{equation}
\begin{aligned}
\langle q \rangle \sim t^{\beta} = t^{-1/2} \\
\sqrt{\sum{|a_m|^2}} \sim t^{\gamma/2} = t^{1/2}.
\label{eq:PowerLawPhi}
\end{aligned}
\end{equation} 

\subsection{Numerical verification of the second to first order reduction}
\begin{figure*}[t!]
\begin{center}
\includegraphics[width=0.7\linewidth,trim=0cm 0cm 0cm 0cm]{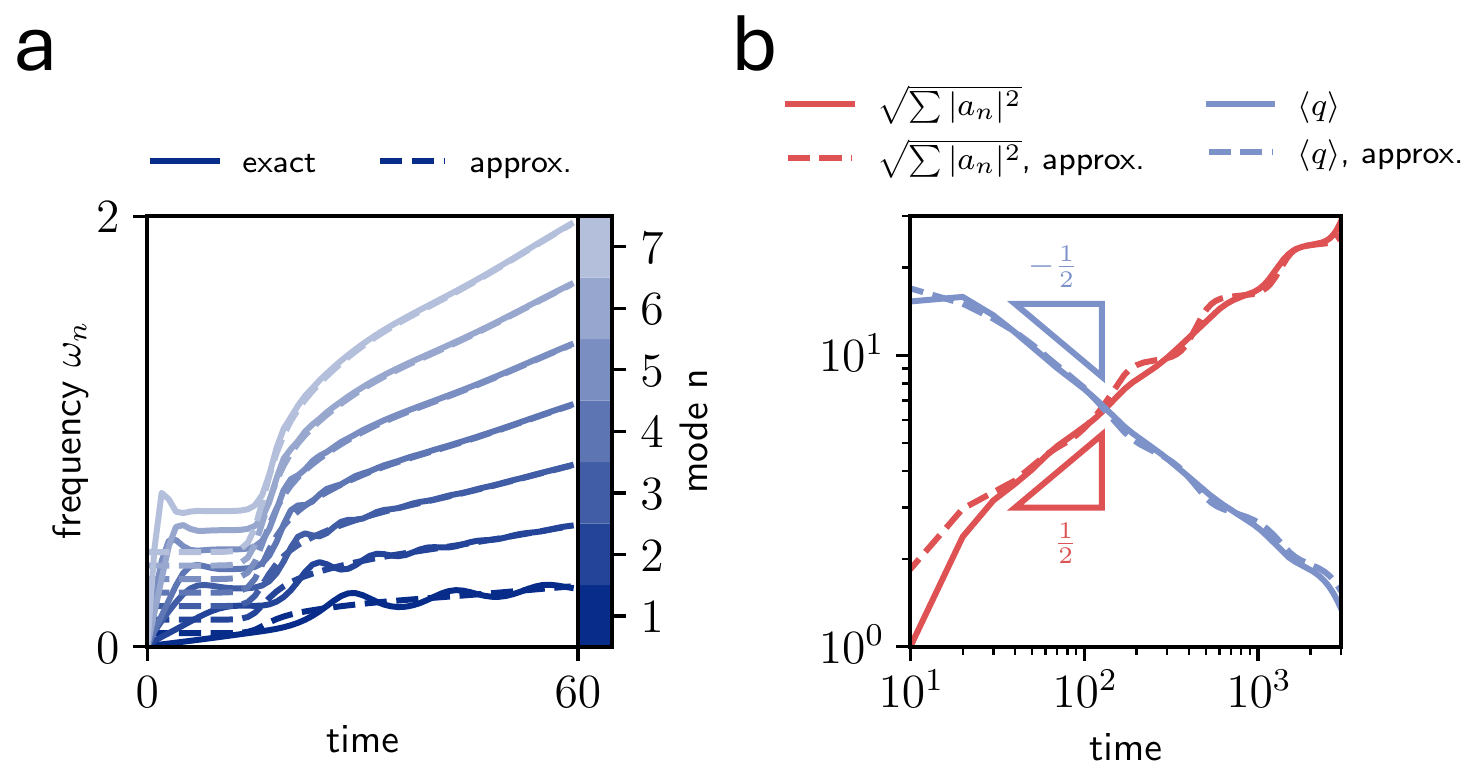}
\end{center}
\caption{\linespread{1.1}\selectfont{}
\textbf{Validation of the multiple scales approximation.} We compare the second order dynamics Eq.~\eqref{eq:k-space-meanfield} with the first order dynamics predicted from a multiple scales approximation. In each plot, solid curves are predictions from the second order dynamics, dashed curves are predictions from the first order dynamics. {\bf (a)} The frequency slaving approximation Eq.~\eqref{eq:NonlinearFrequency} is valid after an initial transient. We show $\omega_n$ for the first $n=8$ modes of a second order simulation (solid), compared to the prediction $\omega_n \sim \sqrt{\Sigma|a_n|^2}$ (dashed). {\bf (b)} The simplified first order dynamics of Eq.~\eqref{eq:FirstOrder} predicts coarsening exponents $\langle q \rangle \sim t^{-1/2}$, $\sqrt{\sum |a_n|^2} \sim t^{1/2}$. We find good agreement between these predictions, and the second order dynamics of Eq.~\eqref{eq:k-space-meanfield}. Here $\alpha=3$ and we simulate the modal dynamics of $n=20$ modes. We stimulate a right-moving wave with initial conditions $a_n \sim N(0,1)$ for $q>0$, $a_n=0$ for $q\leq0$. }
\label{fig:MultipleScales}
\end{figure*}
We make several assumptions in deriving Eq.~\eqref{eq:FirstOrder} from Eq.~\eqref{eq:k-space-meanfield}, which we now numerically verify. In Fig.~\ref{fig:MultipleScales} we compare the simulated dynamics of Eq.~\eqref{eq:k-space-meanfield} with $n=20$ modes to the simplified dynamics and coarsening exponents predicted in Eqs.~\eqref{eq:NonlinearFrequency},~\eqref{eq:FirstOrder} and~\eqref{eq:PowerLawPhi}. In Fig.~\ref{fig:MultipleScales}a we plot instantaneous frequency $\omega_n(t):= \dot{\theta}_n(t)$ compared to the frequency slaving prediction $\omega_n\sim \sqrt{\Sigma |a_m|^2}$ given by Eq.~\eqref{eq:NonlinearFrequency}. We find excellent agreement after the initial linear instability. Eq.~\eqref{eq:PowerLawPhi} further predicts power law decay of $\langle q \rangle \sim t^{-1/2}$ and mean amplitude growth of $\sqrt{\Sigma|a_m|^2} \sim t^{1/2}$. In Fig.~\ref{fig:MultipleScales}b we plot these quantities for the second order dynamics Eq.~\eqref{eq:k-space-meanfield} and the first order dynamics Eq.~\eqref{eq:FirstOrder}. After the initial linear instability we find good agreement with our predicted exponents. 

\section{The nonreciprocal beam and its coarsening exponents}\label{COARSENING:fibre}
In this section we apply the analysis performed above on the $\phi$ model to a non-reciprocal beam theory of the ring pictured in Fig.~\ref{COARSENING:Fig3}, with the continuum model given by Eq.~\eqref{eq:beam}. We begin with the linearized dynamics of a non-reciprocal beam derived in Ref.~\cite{Veenstra_Nature2025}: 
\begin{align}
    m \partial_t^2 h + a^2 \kappa \partial_x^4 h - 2 a^3 \kappa^a \partial_x^5 h + a^2 \gamma  \partial_x^4 \partial_t h =0 \label{eq:beamLINDIM},
\end{align}
where $h$ describes the transverse deflection of the vertices, which have mass $m$, are spaced apart by distance $a$ and are connected with torsional bonds with stiffness $\kappa$, viscous damping $\gamma$ and non-reciprocal interactions $\kappa^a$. The sign of of $\kappa^a$ in Eq.~\eqref{eq:beamLINDIM} is chosen to amplify right-moving waves.

Non-dimensionalizing this equation yields
\begin{align}
\ddot{h} = -\partial^2_x \left[ \partial^2_x h-\alpha \partial^3_x h+ \partial^2_x \dot{h}\right]
, \quad  0\leq x \leq L,
\label{eq:beamLINNONDIM}
\end{align}
with $\alpha = (\kappa^a/\kappa)(\kappa m)^{1/4}\gamma^{-1/2}a^{1/2}$ the dimensionless non-reciprocity and $L$ the ring size. 
To mimic the saturation of the torque motors embedded in the metamaterial of Fig.~\ref{COARSENING:Fig3}, we 
replace the linear active term $\propto \partial_\lambda^5 h$ by a saturating nonlinear function that breaks spatial inversion symmetry whilst retaining the Laplacian form of Eq.~\eqref{eq:beamLINNONDIM}.
We opt for the following substitution:
\begin{equation}
    \partial_x^5 h \rightarrow \partial_x^2\Bigg(\frac{\partial_x^3h}{\sqrt{1+(\partial_x^3 h)^2}}\Bigg),
\end{equation}
and find Eq.~\eqref{eq:beam}:
\begin{equation}
\ddot{h} = -\partial^2_x \left[ \partial^2_x h-\alpha\Bigg(\frac{\partial^3_x h}{\sqrt{1+(\partial^3_x h)^2}}\Bigg)+\partial^2_x \dot{h}\right].
\label{eq:BeamSI}
\end{equation}
\subsection{Mean field and coarsening of the beam model}\label{COARSENING:fibre}
We now perform a series of simplifications to the beam model Eq.~\eqref{eq:BeamSI} which are identical to those performed on the $\phi$ model Eq.~\eqref{eq:phi_modelSI} in \S\ref{COARSENING:AmplitudeEquations}. Fourier transforming as $h = \sum_{n} a_{n} e^{i{q_n}x}$ we obtain 
\begin{align}
    \ddot a_n = -q_n^4\,a_n -q_n^4 \dot{a}_n - \alpha\,q_n^2 \widehat{ \left( \frac{ \partial_{x}^3 h}{\sqrt{1+ ( \partial_{x}^3 h )^2}} \right) }\Bigl._{\!n} 
\end{align}
where the Fourier transform $\widehat{ \left( \frac{\partial_{x}^3 h}{\sqrt{1+ (\partial_{x}^3 h ) ^2}} \right) }\Bigl._{\!n}$ equals $\int^{ L/{2} }_{ -L/2 } \frac{ \partial_{x}^3 h }{\sqrt{1+ (\partial_{x}^3 h) ^2}} e^{i{q_n}x} dx $. We want to obtain only the phase-independent terms, which we can do before the integration over $x$. For a real function $h$ we have $ (\partial_{x}^3 h)^{2} = ( \sum_{l} (i{q_l})^{3} a_{l} e^{i{q_l}x} + \textrm{c.c.}) (\sum_{m} (i{q_m})^{3} a_{m} e^{i {q_m} x} + \textrm{c.c.}) = 2 \sum_{m} q_m^6 a_{m} a_{-m} + \text{phase dependent terms}$. 
We insert this sum over modes into the Fourier transform of the nonlinear term and obtain
\begin{align}
    \ddot{a}_{n} = - q_n^{4} a_{n} - q_n^{4} \dot{a}_{n} + \alpha i q_n^{5}\frac{a_{n}}{\sqrt{1 + 2 \sum_{m} {q^{6}_m} |a_{m}|^{2}}}.
    \label{eq:PhaseIndependentBeam}
\end{align}
We now choose to write $a_{n} = e^{\sigma_{n}}$, as in \S\ref{COARSENING:AmplitudeEquations}. We insert this transformation into Eq.~\eqref{eq:PhaseIndependentBeam} and multiply with $e^{-\sigma_{n}}$ to obtain
\begin{equation}
    \dot{\sigma}_{n}^{2} + \ddot{\sigma}_{n} = -q_n^{4} \left( 1 + \dot{\sigma}_{n} - i q_n \frac{ \alpha }{\sqrt{1 + 2 \sum_{m} q_m^{6} |a_{m}|^{2}}} \right). 
\end{equation}
We now repeat the logic described in \S\ref{COARSENING:AmplitudeEquations} for the $\phi$ model by writing $\dot{\sigma}_{n} = \dot{ \langle \sigma \rangle }_{n} + \dot{\delta}_{n}$. We obtain an expression that is analogous to Eq.~\eqref{eq:dotsigma},
\begin{align}
    \dot{ \sigma }_{n} = - \frac{q_n^{4}}{2} + \frac{q_n^{2}}{2} \sqrt{q_n^{4} - 4 \left( 1 - i \alpha \frac{q_n}{\sqrt{1 + 2 \sum_{m} q_m^{6} |a_{m}|^{2}}} \right)}.
\end{align}
We again are interested in the coarsening, highly nonlinear regime. Taking $\sum_{m} q_m^{6} |a_{m}|^2$ to be much larger than $\alpha$ we Taylor expand to obtain
\begin{align}
    \dot{\sigma}_{n} = - \frac{q_n^{4}}{2} + i q_n^{2} + \frac{q_n^{2}}{2} \alpha \frac{q_n }{\sqrt{1 + 2 \sum_{m} q_m^{6} |a_{m}|^{2}}}.
\end{align}
Here the imaginary component gives the instantaneous frequency $\omega_{n}$, such that
\begin{align}
    \omega_{n} &= q_n^{2}.
\end{align}
The real part gives the instantaneous growth rate such that
\begin{align}
    \dot{|a_{n}|} &= \frac{1}{2} \left( \frac{ \alpha}{\sqrt{1 + 2 \sum_{m} q_m^{6} |a_{m}|^{2}}} q_n^{3} - q_n^{4} \right) |a_{n}|, 
\label{eq:Growthrate_Beam}
\end{align}
\noindent which corresponds to
Eq.~\eqref{eq:NonlinearGrowthRateFirstOrder} of the main text for the case where $p =2$. We proceed by obtaining the exponents for the total amplitude $A$ and central wavenumber $Q$. We define the total amplitude as $ A := \sum_{m} 2 |a_{m}|^{2} q_m^{6} $, while we define the $j$-th moment as $ \langle q^{j} \rangle := \sum_{m} q_m^{j} (2|a_{m}|^{2} q_m^{6})/A$. We again assume a concentrated distribution of amplitude, such that the standard deviation $s \ll Q$, such that $\langle q^{j} \rangle \approx Q^{j}$. 
We again perform the same analysis as for the $\phi$ model, except that we now also multiply both sides by $q_l^{6}$, and we obtain
\begin{align}
    \dot{A} = \alpha A^{1/2} Q^{3} - \tfrac{1}{2} A Q^{4}.
\label{eq:Amplitude_wavenumber_beam}
\end{align}
We again insert the ansatz $Q \propto t^{\beta}$ and $A \propto t^{\chi}$. We likewise see that by equating the powers in $t$, Eq.~\eqref{eq:Amplitude_wavenumber_beam} can only hold when $\beta = -1/4$ and $\chi = 1$. We therefore obtain \begin{align}
    \langle q \rangle &\sim t^{\beta} = t^{-1/4}, \\
    \sqrt{ \sum_{m} q_m^6|a_{m}|^{2}  } &\sim t^{\chi/2} = t^{1/2}.
    \label{eq:FibreCoarseningExponents}
\end{align}
The prediction $\langle q \rangle \sim \xi^{-1} \sim t^{-1/4}$ from Eq.~\eqref{eq:FibreCoarseningExponents} is shown in Fig.~\ref{COARSENING:Fig3}e of the Main Text.

\section{Unidirectionally amplified Rayleigh waves in odd elastic media}\label{COARSENING:Rayleigh}
\begin{figure*}[h]
\centering
\includegraphics[width=0.7\linewidth,trim=0cm 0cm 0cm 0cm]{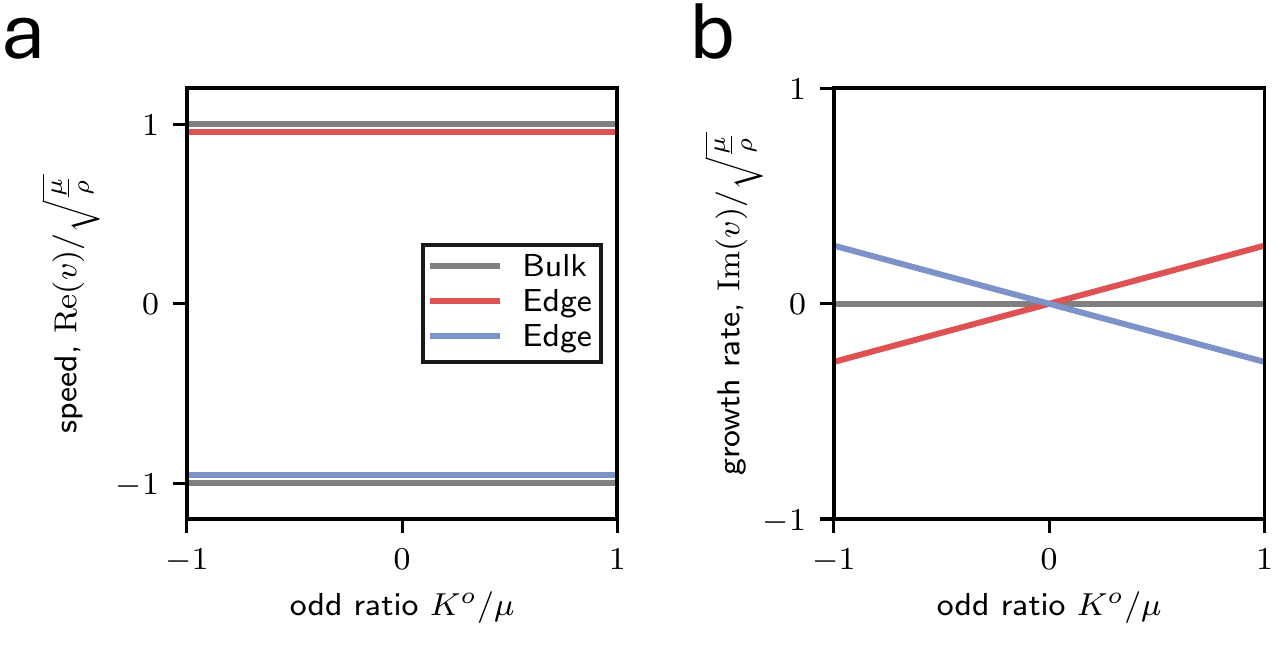} \caption{ \textbf{ Amplification of Rayleigh waves}. Wave speeds $\mathrm{Re}(v)$ and growth rates $\mathrm{Im}(v)$ of edge localized Rayleigh waves, shown as a function of the odd ratio $K^o/\mu$. Here $v=\omega/q_x$ is the velocity of a wave running along a boundary at $y=0$, and we consider an incompressible odd elastic medium, $B \rightarrow \infty$. Note that amplification is unidirectional. For example, a positive $K^o$ amplifies only right-moving waves.}
\label{fig:RayleighWaves}
\end{figure*}
In this section we derive the dispersion relation for Rayleigh waves in an odd elastic medium, and show that they are unidirectionally amplified. Rayleigh waves have been previously been shown to edge localize under open boundaries, with some numerical evidence of their amplification~\cite{Veenstra_Nature2025, chengOddElasticityRealized2021}.

Our starting point is linear odd elastodynamics Eq.~\eqref{eq:LinearDisplacement}. We consider the situation of zero internal torque ($A=0$), zero viscous dissipation $(\eta=0)$ and zero substrate friction ($\Gamma=0)$. Our analysis nevertheless provides insight into a viscously dissipative medium, but only as $q\rightarrow0$. Finite wavenumbers of $O(q^2)$ will undergo viscous damping not considered here, which will stabilize the dispersion at high $q$. We consider a bulk medium occupying the $y>0$ half-plane, with a no stress boundary at $y=0$:
\begin{equation}
\sigma_{xy} = \sigma_{yy}=0, \quad u (y \rightarrow \infty)=0.
\end{equation}
We take a wave ansatz
\begin{equation}
  \begin{bmatrix}
    u_x \\ u_y
  \end{bmatrix}
  =
  \begin{bmatrix}
    \tilde{u}_x \\ \tilde{u}_y
  \end{bmatrix}
  e^{i(q_x x + q_y y - \omega t )}.
\end{equation}
In general, $q_x, q_y, \omega \in \mathbb{C}$. Here we focus on the case $q_x \in \mathbb{R}$, but $\omega, q_y \in \mathbb{C}$, corresponding to a wave propagating along $x$ with complex growth rate $\omega$ and penetration depth $q_y = i \kappa$. We nondimensionalize by $q_x$, defining a dimensionless penetration depth $r= q_y/q_x$ and Rayleigh wave speed $v = \omega/q_x$. Our bulk dispersion relation then reads
\begin{equation}
\begin{bmatrix}
  B + (1+ r^2)\mu - \rho {v}^2 & r B + K^o(1 + r^2) \\
  r B - K^o(1 + r^2) &  B r^2 + (1 + r^2)\mu  - \rho v^2\\
\end{bmatrix}
\begin{bmatrix}
 \tilde{u}_x \\ \tilde{u}_y 
\end{bmatrix}
=0.
\label{eq:BulkDispersion}
\end{equation}
We now require the determinant of Eq.~\eqref{eq:BulkDispersion} to vanish, giving us a secular equation
\begin{equation}
    B \mu +B \mu  r^4+2 B \mu  r^2-B \rho  r^2 v^2-B \rho  v^2+{K^o}^2 r^4+2 {K^o}^2 r^2+{K^o}^2+\mu ^2+\mu ^2 r^4+2 \mu ^2 r^2-2 \mu  \rho  r^2 v^2+\rho ^2 v^4-2 \mu  \rho  v^2
   =0,
\end{equation}
with roots
\begin{equation}
r^{ab} = a \sqrt{\frac{\rho  {v}^2 \left(b \sqrt{B^2-4 {K^o}^2}+B+2 \mu \right)}{2 \left(\mu  (B+\mu )+{K^o}^2\right)}-1},
\label{eq:r}
\end{equation}
where $a,b = \pm 1$ index the four possible roots. Each root comes with a bulk wave eigenvector
\begin{equation}
  n^{ab} = 
\begin{bmatrix}
 \tilde{u}_x \\ \tilde{u}_y 
\end{bmatrix}.
\label{eq:candidate}
\end{equation}
We now combine these eigenvectors to construct Rayleigh waves that satisfy the no-stress boundary conditions. In Fourier space, these boundary conditions read
\begin{align}
   \mu(\tilde{u}_y + r \tilde{u}_x)  +  K^o(r \tilde{u}_y -  \tilde{u}_x) = 0, \\
   B \left(\tilde{u}_x + r \tilde{u}_y \right) - \mu(\tilde{u}_x - r \tilde{u}_y)  - K^o(\tilde{u}_y + r \tilde{u}_x) = 0.
\end{align}
Given a displacement vector $\tilde{u}$, this matching condition can be written as a linear map
\begin{align}
  \begin{bmatrix}
  S_{xy}(\tilde{u}) \\
  S_{yy}(\tilde{u})
  \end{bmatrix}
  =0.
 \label{eq:StressMatch}
\end{align}
We now consider this displacement $\tilde{u}$ to be a linear mix of two candidate Rayleigh wave eigenvectors Eq.~\eqref{eq:candidate} which come from solving the bulk dispersion:
\begin{equation}
  \tilde{u} = p n^{a_1,b_1} + q n^{a_2,b_2},
\end{equation}
with the two modes indexed by the two tuples $(a_1,b_1)$ and $(a_2,b_2)$. Our stress-matching condition Eq.~\eqref{eq:StressMatch} can then be written in matrix form
\begin{align}
S=
  \begin{bmatrix}
  S_{xy}(n^{a_1,b_1}) & S_{xy}(n^{a_2,b_2})  \\
  S_{yy}(n^{a_1,b_1}) & S_{yy}(n^{a_2,b_2})  \\
  \end{bmatrix}
  \begin{bmatrix}
  p \\
  q 
  \end{bmatrix}
  =0,
\end{align}
where each of the $S(n)$ are now known functions. The $\binom{4}{2}=6$ different tuple pairs $(a_1, b_1), (a_2, b_2)$ encode all the possible stress matches coming from mixing eigenvectors. We require the determinant of $S$ to vanish, 
\begin{equation}
   \mathrm{Det}[S(v)] =0
   \label{eq:Solvability}.
\end{equation}
The solvability condition Eq.~\eqref{eq:Solvability} gives an algebraic equation in $v$. We simplify this expression by considering the incompressible limit $B \rightarrow \infty$. Inspection of the resulting polynomial reveals the Hermitian Rayleigh wave to be captured by $(a,b)=(+1,+1), (1,-1)$. With this choice Eq.~\eqref{eq:Solvability} reads
\begin{equation}
{K^o}^2 \left(4+4 i \sqrt{s^2-1}\right)-2 {K^o} \mu  s^2 \left(\sqrt{s^2-1}+i\right)+\mu ^2 \left(s^4-4 s^2+4 i \sqrt{s^2-1}+4\right)=0,
\end{equation}
where $s= v/\sqrt\frac{\mu}{\rho}$ gives the complex Rayleigh wave speed in terms of the shear wave speed. To lowest order in $K^o/\mu$ we find that the solutions to this polynomial which have positive penetration depths are
\begin{equation}
s_\pm = \pm \left[s_r+i \frac{ {K^o}}{\mu }s_i + O\left(\left(\frac{ {K^o}}{\mu }\right)^2\right) \right],
\end{equation}
where $s_r,s_i$ are numerical approximations to exact radical expressions. Here $s_r = 0.955\hdots$ is the standard speed of an incompressible Rayleigh wave. By contrast, $s_i=0.269\hdots$ gives the imaginary part of $s$, and represents wave amplification. Note that only the phase velocity $s_+$ with the same sign as $K^o$ is amplified, with $s_-$ attenuated (Fig.~\ref{fig:RayleighWaves}).

Using Eq.~\eqref{eq:r}, we find the decay lengths of these roots to order $K^o/\mu$ are
\begin{align}
\kappa= 1, \quad  \kappa=  \kappa_r \pm i\frac{ {K^o}}{\mu} \kappa_i,
\end{align}
 in terms of the dimensionless penetration depth $\kappa = -i r$. The real parts of $\kappa$, $\mathrm{Re}(\kappa)= 1,\  \kappa_r=0.296 \hdots$ give the penetration depth of an incompressible Rayleigh wave in a passive medium. In active media we find that these penetration depths also acquire a spatially oscillatory correction $\kappa_i=0.871\hdots$. We conclude that Rayleigh wave states in elastic media continue to exist in the odd elastic context, with the first order correction of non-reciprocity being amplification of these waves in time. Moreover this amplification is thresholdless: even for values of $K^o$ at which bulk modes are stable, surface modes are still amplified. We observe these surface-bound states in the coarsening dynamics shown in Fig.~\ref{COARSENING:Fig4} of the Main Text.
\end{document}